\documentclass[12pt]{article}
\pdfoutput=1
\usepackage{amsmath,amssymb,amsthm,bm,graphicx,float,array,multirow,multicol,rotfloat,caption,subcaption,hyperref,cleveref,enumerate,geometry,mathdots,adjustbox,booktabs,parskip,mathtools,tikz,pdflscape,csquotes,lscape,rotating,empheq}
\usepackage[para]{threeparttable}
\usepackage[all]{xy}
\usepackage[normalem]{ulem}
\usepackage[numbers,sort&compress]{natbib}
\numberwithin{equation}{section}
\numberwithin{figure}{section}
\allowdisplaybreaks
\makeatletter

\usetikzlibrary{arrows,shapes.misc}
\newcommand{\mymk}[1]{%
  \tikz[baseline=(char.base)]\node[anchor=south west, draw,rounded rectangle, inner sep=3pt, minimum size=5mm, text height=3mm](char){\ensuremath{#1}} ;}

\theoremstyle{plain}
\newtheorem*{thm*}{Theorem}

\theoremstyle{definition}

\newtheorem*{defn*}{Definition}

\makeatother

\begin{document}

\begin{titlepage}
\vspace*{-3cm} 
\begin{flushright}
{\tt CALT-TH-2021-026}\\
\end{flushright}
\begin{center}
\vspace{2cm}
{\LARGE\bfseries 
Infinitely many 4d $\mathcal{N}=2$ SCFTs with $a = c$ and beyond
\\}
\vspace{1.2cm}

{\large
Monica Jinwoo Kang$^{1,2}$, Craig Lawrie$^3$, and Jaewon Song$^2$\\}
\vspace{.7cm}
{ $^1$ Walter Burke Institute for Theoretical Physics, California Institute of Technology}\par
{Pasadena, CA 91125, U.S.A.}\par
\vspace{.2cm}
{ $^2$ Department of Physics, Korea Advanced Institute of Science and Technology}\par
{Daejeon 34141, Republic of Korea}\par
\vspace{.2cm}
{ $^3$ Department of Physics and Astronomy, University of Pennsylvania}\par
{Philadelphia, PA 19104, U.S.A.}\par
\vspace{.2cm}

\vspace{.3cm}

\scalebox{.95}{\tt monica@caltech.edu, craig.lawrie1729@gmail.com, jaewon.song@kaist.ac.kr}\par
\vspace{1.2cm}
\textbf{Abstract}
\end{center}

\noindent We study a set of four-dimensional $\mathcal{N}=2$ superconformal field theories (SCFTs) $\widehat{\Gamma}(G)$ labeled by a pair of simply-laced Lie groups $\Gamma$ and $G$. They are constructed out of gauging a number of $\mathcal{D}_p(G)$ and $(G, G)$ conformal matter SCFTs; therefore they do not have Lagrangian descriptions in general. For $\Gamma = D_4, E_6, E_7, E_8$ and some special choices of $G$, the resulting theories have identical central charges $(a=c)$ without taking any large $N$ limit. Moreover, we find that the Schur indices for such theories can be written in terms of that of $\mathcal{N}=4$ super Yang--Mills theory upon rescaling fugacities. Especially, we find that the Schur index of $\widehat{D}_4(SU(N))$ theory for $N$ odd is written in terms of MacMahon's generalized sum-of-divisor function, which is quasi-modular. For generic choices of $\Gamma$ and $G$, it can be regarded as a generalization of the affine quiver gauge theory obtained from $D3$-branes probing an ALE singularity of type $\Gamma$. We also comment on a tantalizing connection regarding the theories labeled by $\Gamma$ in the Deligne--Cvitanovi\'c exceptional series. 

\vfill 
\end{titlepage}

\tableofcontents

\vspace{4cm}

\newpage

\section{Introduction}

It has been a constant desire in modern theoretical physics to unveil the non-perturbative nature of quantum field theories. A natural and rich arena for exploring the non-perturbative aspects of quantum field theory is the realm of four-dimensional superconformal field theories (SCFTs). One of the most striking discoveries in theoretical physics in the past few decades is the understanding of non-conventional quantum field theories, commonly referred as non-Lagrangian field theories \cite{Argyres:1995jj, Argyres:1995xn, Minahan:1996fg, Minahan:1996cj, Argyres:2007cn, Gaiotto:2009we}. They arise in various manners, such as considering the strong-coupling dynamics of string theory and its compactifications, taking infinite-coupling limits of ordinary gauge theories, etc. For instance, the strong-coupling regime of four-dimensional theories is often studied via higher-dimensional origins utilizing geometric and algebraic perspectives.

In particular, 4d $\mathcal{N}=2$ SCFTs have been meticulously analyzed from a string-theoretic perspective. In this paper, we present an interesting class of 4d $\mathcal{N}=2$ SCFTs labeled by two simply-laced Lie groups that we call $\widehat{\Gamma}(G)$, where both $\Gamma$ and $G$ are of type ADE. They are in general non-Lagrangian theories, but for some choices of $\Gamma$ and $G$, the $\widehat{\Gamma}(G)$ theories reduce to familiar Lagrangian quiver gauge theories. For example, when $\Gamma = A_{n-1}$ and $G=SU(N)$, it is simply given by the circular quiver gauge theory with the gauge group $SU(N)^n$ and bifundamental matters that describes the worldvolume theory on $N$ D3-branes probing an $A_{n-1}$ singularity \cite{Douglas:1996sw}.
In general, the $\widehat{\Gamma}(G)$ theory does not necessarily have a weakly-coupled gauge theory description, and it follows that the $\widehat{\Gamma}(G)$ theory is generally non-Lagrangian. 

The superconformal $\widehat{\Gamma}(G)$ theory is constructed via gauging (a subgroup of) the flavor symmetry of a product of superconformal theories. We focus on a particular set of (generally non-Lagrangian) SCFTs known as the $\mathcal{D}_p(G)$ theories \cite{Cecotti:2012jx, Cecotti:2013lda}.\footnote{These theories are identical to the $(G^{(h^\vee)}[p-h^\vee], F)$ theories of \cite{Xie:2012hs, Wang:2015mra} (here $h^\vee$ is the dual Coxeter number of $G$), which are obtained by compactifying 6d $\mathcal{N}=(2, 0)$ theory of type $G \in ADE$ on a sphere with an irregular puncture (labeled by $p$) and a full puncture (F).}\footnote{A similar idea of gauging $\mathcal{D}_p(G)$ theories has been considered in \cite{Closset:2020afy}. They studied some aspects of trivalently-gauged theories, some of which are identical to our $\widehat{E}_{6}(SU(N))$ and $\widehat{E}_{7,8}(G)$ theories.}
The most important feature of the $\mathcal{D}_p(G)$ theory is that it has a flavor symmetry $G$ that can be coupled to a gauge field.\footnote{In general, the flavor symmetry for the $\mathcal{D}_p(G)$ can be enhanced to a larger group than $G$.}
We also consider the conformal matter theories that have a flavor symmetry $G \times G$ \cite{DelZotto:2014hpa}. These conformal matter theories can be obtained by a torus compactification of the 6d $\mathcal{N}=(1, 0)$ conformal matter SCFT with flavor symmetry $G\times G$. This is also identical to the 4d theory obtained from compactifying 6d $\mathcal{N}=(2, 0)$ theory of type $G$ on a sphere with two full punctures and one minimal puncture \cite{Ohmori:2015pua, Ohmori:2015pia}.

We find a particularly interesting feature for the theories with
\begin{align}
    \Gamma = D_4, E_6, E_7, E_8
    \label{eq:gammas}
\end{align}
and for some choices of $G$. More precisely, when the dual Coxeter number $h_G^\vee$ of $G$ and the largest comark $\alpha_\Gamma$ associated to the affine Dynkin diagram $\widehat{\Gamma}$ are co-prime, we find that the two central charges $a$ and $c$ for the $\widehat{\Gamma}(G)$ are equal: 
\begin{align}
\gcd(h_G^\vee, \alpha_\Gamma) = 1 \quad \Longrightarrow \quad a = c .
\end{align}
The largest comarks for the $\Gamma$ in equation \eqref{eq:gammas} are given by 
\begin{align}
    \alpha_{D_4} = 2,\ \alpha_{E_6} = 3,\ \alpha_{E_7} = 4,\ \alpha_{E_8} = 6.
\end{align}
For example, for the choices of $G=SU(N)$, we find that $\widehat{D}_4(SU(N))$ with $N$ odd or $\widehat{E}_6(SU(N))$ with $N=2, 4, 5, 7, \ldots$ have equal central charges, $a=c$. 

As far as we know, there has been almost no known genuine $\mathcal{N}=2$ superconformal field theories with $a=c$. With a larger supersymmetry such as $\mathcal{N}=3$ or $\mathcal{N}=4$, superconformal symmetry implies $a=c$ \cite{Aharony:2015oyb}, but there is no such restriction for $\mathcal{N}=2$ theories. Besides the $\widehat{\Gamma}(G)$ theories, we are only aware of the $(A_{2m}, D_{2m+2}) = D_{2m+2}^{2m+2}[m+1]$ and a small number of $(E_n, G)$ Argyres--Douglas theories for some special choices of $n$ and $G \in ADE$ (in the notation of \cite{Cecotti:2010fi, Wang:2015mra}) that have the same $a$ and $c$ central charges.\footnote{This was noticed in \cite{Agarwal:2017roi} and \cite{Carta:2021whq}, for example. We note that the $(A_{2m}, D_{2m+2})$ and the $\widehat{\Gamma}(G)$ do not overlap, except for $(A_2, D_4)= \widehat{E}_6(SU(2))$. We also find that $\widehat{E}_7(SU(3)) = (E_6, A_3)$, and $\widehat{E}_8(SU(5)) = (E_8, A_5)$. We thank Noppadol Mekareeya for informing us about the $(E_n, G)$ Argyres--Douglas theories with $a=c$ listed in Appendix C of \cite{Carta:2021whq}.}
It is well-known that a holographic theory which has a weakly coupled gravity dual in AdS should have $a=c$ in the large $N$ limit, but it is rather scarce to find four-dimensional conformal field theories with $a=c$ even at finite $N$. Most known holographic theories, including the familiar $\mathcal{N}=2$ SCFTs obtained from $N$ D3-branes probing ALE singularities \cite{Douglas:1996sw, Kachru:1998ys}, have their central charges satisfying 
\begin{align}
    a \sim c \sim O(N^2)\quad\text{and}\quad a-c \sim O(N) \,.
\end{align}
Therefore $a=c$ in the large $N$ limit, but the value $a-c$ is of order $N$ and does not vanish for finite $N$.\footnote{There exists $\mathcal{N}=1$ theories where the central charges scale linearly in $N$: $a \sim c \sim O(N)$, so that $a \neq c$ even for large $N$ \cite{Agarwal:2019crm, Agarwal:2020pol}.}
This particular combination of central charges, $(a-c)$, affects higher-derivative corrections in the supergravity action and contributes to the correction of the famous entropy density--viscosity ratio bound \cite{Kovtun:2004de, Buchel:2008vz}.\footnote{This combination of central charges appears in other contexts as well \cite{DiPietro:2014bca, Perlmutter:2015vma}.} 
In fact, we find that the $\widehat{\Gamma}(G)$ theories, when they are not having $a=c$, can have either signs of $(a-c)$, depending on the choice of $\Gamma$ and $G$.

Another interesting aspect of these theories with $a=c$ is that their Schur indices \cite{Gadde:2011ik, Gadde:2011uv} can be written in terms of the Schur index of $\mathcal{N}=4$ super Yang--Mills theory. In fact, this relationship holds beyond $a = c$ whenever the $\widehat{\Gamma}(G)$ theories have no flavor symmetry. More precisely, we find
\begin{align}
    I_{\widehat{\Gamma}(G)}(q) = I^{\mathcal{N}=4}_{G}(q^{\alpha_\Gamma}, q^{\alpha_\Gamma /2 - 1}) \ , 
\end{align}
where $I^{\mathcal{N}=4}_{G}(q, x)$ refers to the Schur index for the $\mathcal{N}=4$ SYM with gauge group $G$ and $x$ denotes the fugacity for the $SU(2)$ subgroup of the $SU(4)_R$ symmetry. This fact reflects a particularly interesting structure of the spectrum, especially in view of the correspondence between SCFTs and vertex operator algebras \cite{Beem:2013sza}. In the correspondence, the Schur sector is described by a vertex operator algebra (VOA) or a chiral algebra of a 2d CFT, and the Schur index is identical to the vacuum character of the associated VOA. This particular relation involving $\mathcal{N}=4$ SYM is recently found and studied in \cite{Buican:2020moo} for the $G=SU(N)$ case. We find that it generalizes even further to general ADE groups. A similar phenomenon between the Schur index of the $\mathcal{D}_2(SU(N))$ theory with odd $N$ and a free hypermutiplet has been discussed in \cite{Xie:2016evu, Song:2017oew, Buican:2017rya} as well. 

We want to note that we discover a surprising connection between the Schur index and a number-theoretic quantity: we find that the Schur index for $\widehat{D}_4(SU(N))$ with $N$ odd (and equivalently that of $\mathcal{N}=4$ SYM) can be written in terms of MacMahon's generalized sum-of-divisor function \cite{macmahon1921divisors}, which is known to be quasi-modular \cite{MR3028756}.\footnote{We want to make a remark that it is on the 100th anniversary of MacMahon's discovery of the generalized sum-of-divisors function and that it is first appearing in the physics literature!}

Let us remark that our theories can be considered as a natural generalization of affine quiver gauge theories. For example, the theories
\begin{equation}\label{eqn:sudora}
    \widehat{D}_4(SU(2N)) \,, \quad     \widehat{E}_6(SU(3N)) \,, \quad     \widehat{E}_7(SU(4N)) \,, \quad\text{and}\quad
    \widehat{E}_8(SU(6N)) \,
\end{equation}
are simply the worldvolume theory on $N$ D3-branes probing $\mathbb{C}^2/\Gamma$ orbifold, where we abuse notation and use $\Gamma$ to refer to both the ADE group and the discrete subgroup of $SU(2)$ \cite{Douglas:1996sw}. 
There also exists a circular quiver $\widehat{A}_{\ell}(SU(N))$, and an affine $\widehat{D}_{\ell+4}(SU(2N))$ Dynkin diagram shaped quiver.
Apart from these cases, theories corresponding to $\Gamma = {A_{\ell}}, {D_{\ell+4}}$ for $\ell>1$ and $G \neq SU(N)$ need another extra ingredient: the conformal matter of type $(G, G)$. We discuss these cases in detail later in Section \ref{sec:gDM}.

The sequence of affine Lie groups that we consider in equation \eqref{eq:gammas} can be extended to the Deligne--Cvitanovi\'c exceptional series of simply-laced Lie groups \cite{MR1378507,MR2418111}:
\begin{equation}\label{eqn:deligne}
    H_0 \,\subset\, H_1\,\subset\, H_2\,\subset\, D_4\,\subset\, E_6\,\subset\, E_7\,\subset\, E_8 \,.
\end{equation}
We have used the common notation of $H_0$ for the trivial Lie group, $H_1$ for $SU(2)$, and $H_2$ for $SU(3)$ for notational convenience. The Deligne--Cvitanovi\'c exceptional series comes up surprisingly often in physics, and a particularly interesting realization is in the rank-one 4d $\mathcal{N} = 2$ SCFTs obtained from a D3-brane probing an F-theory singularity \cite{Sen:1996vd,Banks:1996nj,Dasgupta:1996ij,Minahan:1996fg,Minahan:1996cj}. In fact, this hints that these 4d theories will have their origins from 6d $(1,0)$ SCFTs as well as $(2,0)$ SCFTs, and hence broadens the scope studied in \cite{Baume:2021qho}. 

To each group appearing in the Deligne--Cvitanovi\'c exceptional series, there is an associated rational number 
\begin{align}
    \Delta_\Gamma=1+\frac{h_\Gamma^\vee}{6},
\end{align}
which we list in Table \ref{tbl:deltas}. This $\Delta_\Gamma$ corresponds to the scaling dimension of the Coulomb branch operator of the corresponding rank-one $\mathcal{N}=2$ superconformal theory. The central charge $c$ of the four families $\widehat{\Gamma}(G)$ of theories with $\Gamma$ as in equation \eqref{eq:gammas} can be written as
\begin{equation}\label{eqn:lrgNc}
    c \sim \frac{\Delta_\Gamma - 1}{\Delta_\Gamma} \text{dim}(G) \,.
\end{equation}
This expression for $c$ suggests the existence of theories with the same dependence of the central charge $c$ to the choice of $\Gamma$ and $G$ as in equation \eqref{eqn:lrgNc} in the Deligne--Cvitanovi\'c exceptional series: namely,
\begin{equation}
    \widehat{H}_0(G) \,, \quad \widehat{H}_1(G) \,, \quad
    \widehat{H}_2(G) \,.
\end{equation}
In fact, we find the corresponding theories for $\widehat{H}_0(G)$ and $\widehat{H}_1(G)$ are a gauge node with no matter and the $\mathcal{N}=4$ super Yang--Mills theory, respectively. We further propose a candidate theory corresponding to the $\widehat{H}_2(G)$, when $G$ is a classical group, as a gauge node $G$ coupled to the correct number of fundamental hypermultiplets to cancel the one-loop $\beta$-function.

\begin{table}[H]
    \begin{threeparttable}
    \centering
    \renewcommand{\arraystretch}{1.2}
    $\begin{array}{c|ccccccc}
    \toprule
         \Gamma & H_0 & H_1 & H_2 & D_4 & E_6 & E_7 & E_8 \\\midrule
         \Delta_\Gamma & \frac{6}{5} & \frac{4}{3} & \frac{3}{2} & 2 & 3 & 4 & 6 \\
    \bottomrule
    \end{array}$
    \end{threeparttable}
    \caption{The rational invariants $\Delta_\Gamma$ associated to each entry, $\Gamma$, in the Deligne--Cvitanovi\'c exceptional series given in equation \eqref{eqn:deligne}.}
    \label{tbl:deltas}
\end{table}

The organization of the remainder of the paper is as follows. In Section \ref{sec:gauge}, we first review the theories $\mathcal{D}_p(G)$ and discuss the limited number of ways these theories can be gauged together, which yields the SCFTs $\widehat{\Gamma}(G)$. In Section \ref{sec:prop}, we determine the central charges, the scaling dimensions of the Coulomb branch operators, and other physical information about the $\widehat{\Gamma}(G)$. We then explore in Section \ref{sec:aequalsc} in more detail the $\widehat{\Gamma}(G)$ which have $a=c$, and we determine the Schur index of these theories in Section \ref{sec:index}. We then explain how the Schur indices can be written as the Schur index of $\mathcal{N}=4$ super Yang--Mills with rescaled fugacities in Section \ref{sec:N4index}. In Section \ref{sec:ext1}, we explore some of the theories with $a \neq c$, and we discuss a generalization where $\Gamma$ takes values in the simply-laced Deligne--Cvitanovi\'c series of Lie groups. In Section \ref{sec:gDM}, we discuss the extension to $\Gamma$ being an arbitrary ADE group, which gives a generalization of all the affine quivers. Finally, we conclude with possible future directions in Section \ref{sec:discussion}.

\section{Constructing $\widehat{\Gamma}(G)$ SCFTs}\label{sec:two}

We construct the superconformal theories $\widehat{\Gamma}(G)$ by gluing (multiple) copies of the $\mathcal{D}_p(G)$ theories, which we introduce and explain in Section \ref{sec:gauge} and further explore their properties and physical significance in Section \ref{sec:prop}.

\subsection{Gauging $\mathcal{D}_p(G)$s}\label{sec:gauge}

In this section, we perform the construction of the superconformal theories that we call $\widehat{\Gamma}(G)$. The $\widehat{\Gamma}(G)$ theory is built out of gluing copies of the $\mathcal{D}_p(G)$ SCFTs; these theories were introduced in \cite{Cecotti:2012jx} and explored further in \cite{Cecotti:2013lda}.\footnote{We denote this theory using the calligraphic $\mathcal{D}$ instead of $D$ to avoid any possible confusion with the Lie group of $D$-type.} 
The $\mathcal{D}_p(G)$ theory is labeled by a simply-laced Lie group $G \in ADE$ and a positive integer $p$. This theory has a flavor symmetry that is at least $G$. In the class $\mathcal{S}$ framework, the theory $\mathcal{D}_p(G)$ can also be constructed as a compactification of a 6d $(2,0)$ SCFT on a sphere with one regular and one irregular puncture. In that construction, they are written as $(G^b[p - h_G^\vee], F)$ \cite{Wang:2015mra}, where $b = h_G^\vee$ and $F$ denotes a full puncture. From this perspective, the flavor symmetry $G$ arises from the full puncture and any extra or enhanced symmetry is due to the irregular puncture that is labeled by $p$. See Table \ref{tbl:DpGnoextraF} for the condition for $\mathcal{D}_p(G)$ not to have any enhanced symmetry besides $G$. 

\begin{table}[H]
    \begin{threeparttable}
    \centering
    \renewcommand{\arraystretch}{1.2}
    \centering
    $\begin{array}{c|ccccc}
        \toprule
        G & SU(N) & SO(2N) & E_6 & E_7 & E_8 \\
        \midrule
        \text{No additional symmetry} & (p, N) = 1 & p \notin 2\mathbb{Z}_{>0} & p \notin 3\mathbb{Z}_{>0} & p \notin 2\mathbb{Z}_{>0} & p \notin 30\mathbb{Z}_{>0}
        \\\bottomrule
    \end{array}$
    \end{threeparttable}
    \caption{The condition for $\mathcal{D}_p(G)$ to have no extra symmetry besides $G$. Equivalently, the condition for the irregular puncture not to carry any flavor symmetry.}
    \label{tbl:DpGnoextraF}
\end{table}

Let us explore the ways in which a collection of theories $\mathcal{D}_{p_i}(G)$ can be gauged together by their common flavor symmetry $G$. 
In order to obtain $\mathcal{N}=2$ superconformal theory upon gauging, we require the beta function for the gauge coupling to vanish, which turns out to be highly constraining. 

To show that the conformal gauging of the common $G$ flavor symmetry is restrictive, we begin by gauging together the $G$ of $\mathcal{D}_{p_i}(G)$ for $i = 1, \cdots, n$. The conformal gauging condition is given by  
\begin{equation}
\label{eqn:confgauge}
    \sum_{i=1}^n k_i = 4 h_G^\vee \,.
\end{equation}
where $h_G^\vee$ is the dual Coxeter number of the gauge group $G$. The dual Coxeter numbers, together with the other relevant group-theoretic quantities, are listed in Table \ref{tbl:gpdata}.
The flavor central charge $k_i$ for each of the $\mathcal{D}_{p_i}(G)$ is determined as \cite{Cecotti:2013lda, Xie:2012hs}
\begin{equation}
    k_i = \frac{2(p_i - 1)}{p_i}h_G^\vee \,.
\end{equation}
The conformal gauging condition \eqref{eqn:confgauge} is then simply the constraint 
\begin{equation}\label{eqn:confgauge2}
    \sum_{i = 1}^n \frac{1}{p_i} = n - 2 \,.
\end{equation}
The $\mathcal{D}_p(G)$ theories are associated to integers $p > 0$, and the theory $\mathcal{D}_1(G)$ is taken to represent the trivial theory. It follows that $n\geq 2$ as the left hand side of the equation \eqref{eqn:confgauge2} is non-negative. Since if $(p_1,\cdots,p_n)$ is a solution then $(p_1,\cdots,p_n,p_{n+1}=1)$ is also a solution, we restrict no more than one $p_i$ to be one  for $n\geq 4$ to avoid repetitions. (This allows $n=3$ solutions to be accounted as a solution with $n=4$ with one $p_i$ as one.) It is straightforward to see that equation \eqref{eqn:confgauge2} then does not have any solution for $n>4$ with this restriction. We can see that all possible solutions with finite $p_i$ are
\begin{equation}
    \begin{aligned}
    (p_1,p_2,p_3,p_4)=(2, 2, 2, 2),\ (1, 3, 3, 3),\ (1, 2, 4, 4),\ (1, 2, 3, 6) \,,
    \end{aligned}
    \label{eq:4solutions}
\end{equation}
as was also demonstrated in \cite{Cecotti:2013lda}.
We label these solutions as $\widehat{D}_4$, $\widehat{E}_6$, $\widehat{E}_7$, and $\widehat{E}_8$, respectively. This is because they correspond to the type of quivers forming  respectively $\widehat{D}_4$, $\widehat{E}_6$, $\widehat{E}_7$, and $\widehat{E}_8$, once we consider $\mathcal{D}_p(G)$ theory that has a Lagrangian description. We will explain and discuss this point later in this section.

There are also solutions where some of the $p_i$ are infinite:
\begin{equation}
    \begin{aligned}
    (p_1,p_2,p_3,p_4)=(1, 2, 2, \infty),\ (1, 1, \infty, \infty) \,,
    \end{aligned}
\end{equation}
however, it is presently unclear what the theories $\mathcal{D}_\infty(G)$ are in the sense of superconformal field theories. These would be putative theories with a dense Coulomb branch spectrum and a flavor symmetry $G$ with level $2h_G^\vee$. It would be intriguing to explore the potential existence of such theories, which we leave for future work.

\begin{table}[H]
    \begin{threeparttable}
    \centering
    \renewcommand{\arraystretch}{1.1}
    $\begin{array}{cccccc}
    \toprule
        G & d_G & r_G & h_G^\vee & \Lambda_G & \text{Casimir degrees}  \\\midrule
        SU(K) & K^2 - 1 & K - 1 & K & K & 2,3,\cdots,K \\
        SO(2K) & K(2K-1) & K & 2K - 2 & 4K - 8 & 2, 4, \cdots, 2K-2, K \\
        E_6 & 78 & 6 & 12 & 24 & 2,5,6,8,9,12 \\
        E_7 & 133 & 7 & 18 & 48 & 2,6,8,10,12,14,18 \\
        E_8 & 248 & 8 & 30 & 120 & 2,8,12,14,18,20,24,30 \\
    \bottomrule
    \end{array}$
    \end{threeparttable}
    \caption{The relevant data associated to an ADE Lie group $G$. We list for each $G$ the dimension ($d_G$), rank ($r_G$), dual Coxeter number ($h_G^\vee$), the order of the finite ADE subgroup of $SU(2)$ ($\Lambda_G$), and the degrees of the Casimir invariants.}
    \label{tbl:gpdata}
\end{table}

\begin{table}[H]
\centering
\begin{threeparttable}
\small
\renewcommand{\arraystretch}{1.1}
\begin{tabular}{cccc}
\toprule
$(p_1,p_2,p_3,p_4)$ & $\widehat{\Gamma}(G)$ & Quivers via gauging $\mathcal{D}_p(G)$s & $a=c$ \\\midrule
$(2,2,2,2)$ & $\widehat{D}_4(G)$ &
\begin{tikzpicture}[baseline={([yshift=-.3ex]current bounding box.center)}]
\node[anchor=south west](At) at (0.4,1.0) {$\mathcal{D}_{2}(G)$};
\node[anchor=south west](Ab) at (0.4,-1.2) {$\mathcal{D}_{2}(G)$};
\node[anchor=south west](A1) at (-1.1,-0.05) {$\mathcal{D}_{2}(G)$};
\node[anchor=south west, draw,rounded rectangle, inner sep=3pt, minimum size=5mm, text height=3mm](A2) at (1,0) {$G$};
\node[anchor=south west](A3) at (1.85,-0.05) {$\mathcal{D}_{2}(G)$};
\draw (A1)--(A2)--(A3);
\draw (At)--(A2)--(Ab);
\end{tikzpicture}
& $\frac{1}{2}\text{dim}(G) $ \\[6.5ex]
$(1,3,3,3)$ & $\widehat{E}_6(G)$ &
\begin{tikzpicture}[baseline={([yshift=-.3ex]current bounding box.center)}]
\node[anchor=south west](At) at (0.4,1.0) {$\mathcal{D}_{3}(G)$};
\node[anchor=south west](A1) at (-1.1,-0.05) {$\mathcal{D}_{3}(G)$};
\node[anchor=south west, draw,rounded rectangle, inner sep=3pt, minimum size=5mm, text height=3mm](A2) at (1,0) {$G$};
\node[anchor=south west](A3) at (1.85,-0.05) {$\mathcal{D}_{3}(G)$};
\draw (A1)--(A2)--(A3);
\draw (At)--(A2);
\end{tikzpicture}
& $\frac{2}{3}\text{dim}(G) $ \\[4ex]
$(1,2,4,4)$ & $\widehat{E}_7(G)$ &
\begin{tikzpicture}[baseline={([yshift=-.3ex]current bounding box.center)}]
\node[anchor=south west](At) at (0.4,1.0) {$\mathcal{D}_{2}(G)$};
\node[anchor=south west](A1) at (-1.1,-0.05) {$\mathcal{D}_{4}(G)$};
\node[anchor=south west, draw,rounded rectangle, inner sep=3pt, minimum size=5mm, text height=3mm](A2) at (1,0) {$G$};
\node[anchor=south west](A3) at (1.85,-0.05) {$\mathcal{D}_{4}(G)$};
\draw (A1)--(A2)--(A3);
\draw (At)--(A2);
\end{tikzpicture}
& $\frac{3}{4}\text{dim}(G) $ \\[4ex]
$(1,2,3,6)$ & $\widehat{E}_8(G)$ &
\begin{tikzpicture}[baseline={([yshift=-.3ex]current bounding box.center)}]
\node[anchor=south west](At) at (0.4,1.0) {$\mathcal{D}_{2}(G)$};
\node[anchor=south west](A1) at (-1.1,-0.05) {$\mathcal{D}_{3}(G)$};
\node[anchor=south west, draw,rounded rectangle, inner sep=3pt, minimum size=5mm, text height=3mm](A2) at (1,0) {$G$};
\node[anchor=south west](A3) at (1.85,-0.05) {$\mathcal{D}_{6}(G)$};
\draw (A1)--(A2)--(A3);
\draw (At)--(A2);
\end{tikzpicture}
& $\frac{5}{6}\text{dim}(G) $ \\
\bottomrule
\end{tabular}
\end{threeparttable}
\caption{All the solutions with finite $p_i$ and the corresponding $\widehat{\Gamma}(G)$ theories when $a=c$. The theories have $a=c$ when $\gcd(\alpha_\Gamma, h_g^\vee) = 1$, which restricts the $G$ to be those in Table \ref{tbl:allac}.}
\label{tbl:1gaugenode}
\end{table}

The diagonal gauging of the common flavor symmetry $G$ leads to a quiver-like structure
\begin{align}\label{eqn:trivone}
\begin{aligned}
\begin{tikzpicture}
\node(At) at (1.05,1.4) {$\mathcal{D}_{p_1}(G)$};
\node(Ab) at (1.05,-0.9) {$\mathcal{D}_{p_4}(G)$};
\node(A1) at (-0.6,0.25) {$\mathcal{D}_{p_2}(G)$};
\node[anchor=south west, draw,rounded rectangle, inner sep=3pt, minimum size=5mm, text height=3mm](A2) at (1,0) {$G$};
\node(A3) at (2.7,0.25) {$\mathcal{D}_{p_3}(G)$};
\draw (A1)--(A2)--(A3);
\draw (At)--(A2)--(Ab);
\end{tikzpicture}
\end{aligned}
\end{align}
and for the three cases with $p_i=1$, the corresponding node is omitted as $\mathcal{D}_1(G)$ is trivial. For each case in equation \eqref{eq:4solutions}, the gauging is depicted in Table \ref{tbl:1gaugenode}. There are special cases where $\mathcal{D}_p(G)$ is given by a Lagrangian quiver. For instance, by taking $G$ to be
\begin{align}
    G=SU(N)\quad\text{for}\quad N=p\ell,
\end{align}
we can utilize the relation 
\begin{align}\label{eqn:lagexp}
\begin{aligned}
\begin{tikzpicture}
\node at (-1.7,0.3) {$\mathcal{D}_p(SU(p\ell)) =$};
\node[anchor=south west, draw, rectangle, inner sep=3pt, minimum size=5mm, text height=3mm](A0) at (0,0) {$SU(p\ell)$};
\node[anchor=south west, draw,rounded rectangle, inner sep=3pt, minimum size=5mm, text height=3mm](A1) at (2.5,0) {$SU((p-1)\ell)$};
\node(A2) at (6,0.25) {$\cdots$};
\node[anchor=south west, draw,rounded rectangle, inner sep=3pt, minimum size=5mm, text height=3mm](A3) at (7.4,-0.07) {$SU(\ell)$};
\draw (A0)--(A1)--(A2)--(A3);
\end{tikzpicture}
\end{aligned}
\end{align}
which is a Lagrangian theory.
Then we are left to find which choices of $p_i$ and $G$ can make the quiver in equation \eqref{eqn:trivone} to be a Lagrangian quiver.

\begin{figure}[H]
\centering
\vspace{10pt}
\begin{subfigure}[b]{0.45\textwidth}
\centering
\includegraphics{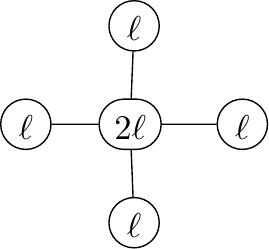}
\caption{The $\widehat{D}_4$ Lagrangian quiver when $(p_1, p_2, p_3, p_4) = (2, 2, 2, 2)$ and $G = SU(2\ell)$.}
\label{fig:d4quiver}
\end{subfigure}\hspace{6mm}
\begin{subfigure}[b]{0.45\textwidth}
\centering
\includegraphics{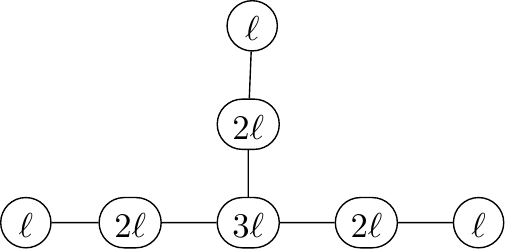}
\caption{The $\widehat{E}_6$ Lagrangian quiver when $(p_1, p_2, p_3, p_4) = (1, 3, 3, 3)$ and $G = SU(3\ell)$.}
\label{fig:e6quiver}
\end{subfigure}\\[16pt]
\begin{subfigure}[b]{0.8\textwidth}
\centering
\includegraphics{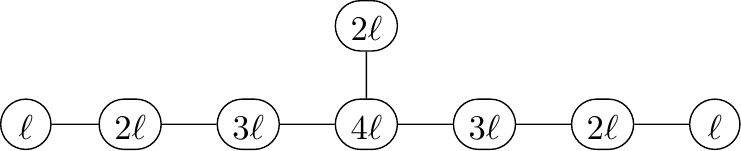}
\caption{The $\widehat{E}_7$ Lagrangian quiver when $(p_1, p_2, p_3, p_4) = (1, 2, 4, 4)$ and $G = SU(4\ell)$.}
\label{fig:e7quiver}
\end{subfigure}\\[16pt]
\begin{subfigure}[b]{0.8\textwidth}
\centering
\includegraphics{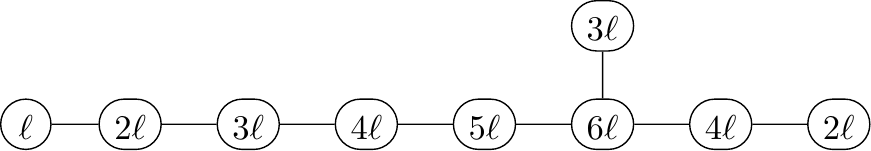}
\caption{The $\widehat{E}_8$ Lagrangian quiver when $(p_1, p_2, p_3, p_4) = (1, 2, 3, 6)$ and $G = SU(6\ell)$.}
\label{fig:e8quiver}
\end{subfigure}
\caption{When the gauge group $G$ appearing in the quiver in equation \eqref{eqn:trivone} is an $SU(N)$ group such that each $p_i$ divides $N$, then one can use the description in equation \eqref{eqn:lagexp} to rewrite \eqref{eqn:trivone} as a Lagrangian quiver. We depict such Lagrangian quivers and observe that these are the standard affine quiver gauge theories that arise on the worldvolume of D3-branes probing $\mathbb{C}^2/\Gamma$ orbifolds \cite{Douglas:1996sw}. Here, we introduce the shorthand notation of writing $N$ inside of a gauge node to represent an $SU(N)$ gauge group.}
\label{fig:affquiv}
\end{figure}

In fact, we find that for the four solutions in equation \eqref{eq:4solutions}, with a particular choice of $G$ to yield a Lagrangian theory, give rise to the known affine $D_4$, $E_6$, $E_7$, and $E_8$ quivers, respectively.  When $(p_1, p_2, p_3, p_4) = (2,2,2,2)$ and $G =SU(2\ell)$, we get the Lagrangian quiver corresponding to $\widehat{D}_4$ as depicted in Figure \ref{fig:d4quiver}. When $(p_1, p_2, p_3, p_4) = (1, 3, 3, 3)$ and $G =SU(3\ell)$, we have Figure \ref{fig:e6quiver}, which shows the Lagrangian quiver of type $\widehat{E}_6$. Similarly, for $(p_1, p_2, p_3, p_4) = (1, 2,4,4)$ and $G = SU(4\ell)$, we find that it gives rise to the Lagrangian quiver of type $\widehat{E}_7$ as Figure \ref{fig:e7quiver}; for $(p_1, p_2, p_3, p_4) = (1, 2,3,6)$ and $G = SU(6\ell)$, we see the Lagrangian quiver of type $\widehat{E}_8$, as represented in Figure \ref{fig:e8quiver}. In each of these cases, these are the Lagrangian quivers known to arise as the worldvolume theory on a stack of $\ell$ D3-branes probing a $\mathbb{C}^2/\Gamma$ orbifold \cite{Douglas:1996sw}. 

\begin{figure}[H]
    \centering
    \begin{subfigure}[b]{0.45\textwidth}
        \centering
        \includegraphics[scale=1.6]{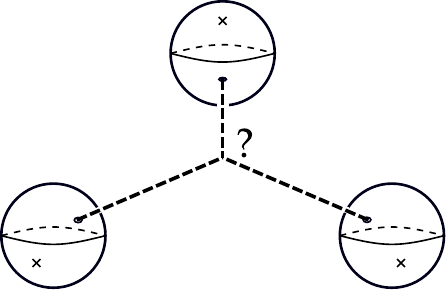}
        \vspace{3mm}
        \caption{The gluing.}
        \label{subfig:triglue}
    \end{subfigure}
    \hspace{6mm}
    \begin{subfigure}[b]{0.45\textwidth}
        \centering
        \includegraphics[scale=2.5]{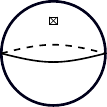}
        \vspace{3mm}
        \caption{The one-punctured sphere.}
        \label{subfig:trionep}    
    \end{subfigure}\\[3pt]
    \caption{The trivalent gauging of the common $G$ flavor symmetry of $\mathcal{D}_{p_1}(G)$, $\mathcal{D}_{p_2}(G)$, and $\mathcal{D}_{p_3}(G)$ is depicted in (a). Each of the three theories has an interpretation as a sphere with a irregular puncture and a regular puncture, where the latter contributes the flavor symmetry, $G$. The regular punctures are glued together to gauge $G$, and the three irregular punctures, denoted by crosses, remain. In known examples, this is equivalent to a one-punctured sphere, depicted in (b), where the puncture (denoted as a boxed cross) is formed by coalescing the three irregular punctures.}
    \label{fig:trigauging}
\end{figure}

We therefore conclude that there are only four combinations of finite $p_i$ for which such a gauging leads to a conformal theory. Using the interpretation of $\mathcal{D}_{p_i}(G)$ as a two-punctured sphere, this gauging involves connecting the regular punctures in a multi-valent vertex. It is important to note that such a gauging is not something that happens inside the class $\mathcal{S}$ construction \cite{Gaiotto:2009hg, Gaiotto:2009we, Bonelli:2011aa, Xie:2012hs, Wang:2015mra} where, for example, a trivalent vertex occurs when one includes a copy of the $T_N$ theory. In the examples where we know the resulting theory after gauging, we can see that the remaining three irregular punctures coalesce into a new kind of irregular puncture, however, we emphasize that we do not observe this behaviour in general. For example, $\widehat{E}_6(SU(2))$ theory is formed by gluing three copies of $\mathcal{D}_3(SU(2))=H_1$ Argyres--Douglas theories. The gauged theory turns out to be identical to the $D_4^6[3]$ Argyres--Douglas theory \cite{Wang:2015mra, Xie:2016evu}. See Figure \ref{fig:trigauging} for the illustration. \\

In turn, we may find the overlap between $\widehat{\Gamma}(G)$ and $J^b[k]$. The theories $J^b[k]$ are obtained by compactifying the 6d $(2,0)$ SCFT of type $J$ on a sphere with a single irregular puncture, where $J$ is an $ADE$ Lie group. The values of $b$ and $k$ specify the choice of irregular solution of Hitchin's equation \cite{hitchin1987stable,biquard2004wild,boalch2012simply,boalch2008irregular,boalch2005klein} that defines an irregular puncture. It has been established in \cite{Agarwal:2016pjo, Giacomelli:2017ckh} that the theories $J^b[k]$ are realized as the endpoint of an $\mathcal{N}=1$ deformation of \cite{Maruyoshi:2016tqk, Maruyoshi:2016aim, Agarwal:2016pjo} (i.e.~an $\mathcal{N}=1$-preserving principal nilpotent deformation followed by a renormalization group flow after which supersymmetry is enhanced back to $\mathcal{N} = 2$) of the SCFTs $\mathcal{D}^b_k(J)$.\footnote{When $b = h_J^\vee$, then we drop $b$ from the notation. These are then the $\mathcal{D}_p(G)$ theories which we have been gauging to obtain the $\widehat{\Gamma}(G)$.} 

\subsection{Properties of $\widehat{\Gamma}(G)$}\label{sec:prop}

In this section, we study the properties of the $\widehat{\Gamma}(G)$ theories. We focus on the cases where 
\begin{align}
    \Gamma = D_4,\ E_6,\ E_7,\ E_8,
\end{align} 
which are typically non-Lagrangian theories. (For specific choices of $G$, the Lagrangian quivers corresponding to these $\widehat{\Gamma}(G)$ are depicted in Figure \ref{fig:affquiv}.) As shown in equation \eqref{eqn:trivone}, these theories can be constructed via gauging the diagonal of the common $G$ flavor group of a set of $\mathcal{D}_{p_i}(G)$ theories. The physical features of the $\widehat{\Gamma}(G)$ can be written in terms of the properties of the constituent $\mathcal{D}_{p_i}(G)$s. In this section, we study the flavor symmetry, the Coulomb branch operator dimensions, and the central charges of these $\widehat{\Gamma}(G)$.

First, we determine the rank of the flavor symmetry of the theories $\widehat{\Gamma}(G)$ by utilizing the knowledge of the flavor symmetries of the $\mathcal{D}_{p_i}(G)$. Since the $\mathcal{D}_p(G)$ theory always has $G$ as (a part of) the flavor symmetry, the rank of the flavor symmetry can be written as
\begin{equation}
    \text{rank}(G) + f(p_i, G) \,,
\end{equation}
where the $f(p_i, G)$ is the `extra symmetry' the $\mathcal{D}_p(G)$ theory has besides $G$ \cite{Cecotti:2013lda}. In terms of the class $\mathcal{S}$ description of Argyres--Douglas theory \cite{Xie:2012hs, Wang:2015mra}, the flavor symmetry $G$ arises from the full puncture (F), and the rest comes from the irregular puncture of type $G^b[p-h_G^\vee]$, with $b=h_G^\vee$. Along with the choice of $G$, the parameter $p$ determines the order of the singularity of the Seiberg--Witten (SW) geometry corresponding to the $\mathcal{D}_p(G)$ theory. When it is possible to have a deformation parameter of the SW curve to be of dimension-one, this parameter corresponds to the mass parameter. When the irregular singularity gives rise to such a mass parameter, we have non-vanishing $f(p, G)$. Considering as an example when $G=SU(N)$, if $p$ and $N$ are co-prime, the irregular singularity is free of such an extra mass parameter so that $f(p, G)=0$; however, if $p$ and $N$ are not co-prime, there exist $\gcd(p, N)-1$ mass parameters associated with the irregular singularity. For the values of $p$ relevant to our interest (i.e.~for $p=2,3,4,6$), we have
\begin{subequations}
\begin{empheq}[left=\empheqlbrace]{align}
    f(p, SU(N)) &= \gcd(p, N) - 1 \,,\\
    f(p_\text{even}, SO(2N)) &= \gcd(p_\text{even}, 2N - 2) - \gcd(p_\text{even}, N-1) + 1 \,,\\
    f(p_\text{odd}, SO(2N)) &= 0 \,,\\
    f(p, E_6) &= \gcd(p, 3) - 1 \,,\\
    f(p, E_7) &= \gcd(p, 2) - 1 \,,\\
    f(p, E_8) &= \gcd(p, 1) - 1 = 0 \,.
\end{empheq}
\label{eqn:fpGall}
\end{subequations}
After the gauging by which we construct the theory $\widehat{\Gamma}(G)$, we find the rank of the remaining flavor symmetry to be simply
\begin{equation}
    f(\widehat{\Gamma}(G)) = \sum_i f(p_i, G) \,.
\end{equation}

The scaling dimensions of the operators that parametrize the Coulomb branch can also be determined from the $\mathcal{D}_{p_i}(G)$ building block theories. We denote by $\mathcal{C}(p_i, G)$ the collection of the Coulomb branch scaling dimensions of the theory $\mathcal{D}_{p_i}(G)$; we further define $\text{Cas}(G)$ to be the set composed of the degrees of the Casimirs of $G$. The Casimir degrees for each $G$ are summarized in Table \ref{tbl:gpdata}. Then, the Coulomb branch operator spectrum $\mathcal{C}(p, G)$ for the $\mathcal{D}_p(G)$ theory can be easily determined from the Seiberg--Witten geometry and can be written as
\begin{align}
    \mathcal{C}(p, G) = \left\{j - \frac{h_G^\vee}{p}s \,\Big|\, j - \frac{h_G^\vee}{p}s > 1\,,\, j \in \text{Cas}(G)\,,\, s = 1, \cdots, p-1 \right\} \,.
\end{align}
The dimension of the Coulomb branch or the rank of $\mathcal{D}_p(G)$ theory is now given as
\begin{align}
    \textrm{rank}(\mathcal{D}_p(G)) = \frac{1}{2} \left( (p-1)\textrm{rank}(G) - f(p, G) \right) \ . 
\end{align}
In terms of these quantities, we find that the Coulomb branch operators of $\widehat{\Gamma}(G)$ have scaling dimensions
\begin{equation}
    \text{Cas}(G) \oplus \bigoplus_{i} \mathcal{C}(p_i, G) \,.
\end{equation}
The $\text{Cas}(G)$ part comes from the gauge group $G$. 
By using the relation between the parameters $p_i$ and the $\Gamma$,
\begin{equation}
    \sum_i (p_i - 1) = \text{rank}(\Gamma) \,,
\end{equation}
we find the rank of the $\widehat{\Gamma}(G)$ theory, which is simply the dimension of the Coulomb branch, to be 
\begin{equation} \label{eq:rankGhatG}
    \text{rank}(\widehat{\Gamma}(G)) =
    \text{rank}(G)\left(1+\frac{\text{rank}(\Gamma)}{2}\right)-\frac{1}{2}f(\widehat{\Gamma}(G))\,.
\end{equation}
For the $\mathcal{D}_p(G)$ theories without any extra global symmetry besides $G$, we have $f(p, G)=0$. When we form $\widehat{\Gamma}(G)$ theory from such theories, the last term in \eqref{eq:rankGhatG} vanishes. 

Finally, we turn our attention to determining the central charges of $\widehat{\Gamma}(G)$. The central charges of each individual $\mathcal{D}_{p_i}(G)$, which we write as $a(p_i, G)$ and $c(p_i, G)$, are somewhat intricate expressions \cite{Shapere:2008zf, Xie:2012hs, Cecotti:2013lda}; however, we observe a pleasant simplification on these central charges when combined to express the central charges for the $\widehat{\Gamma}(G)$ theories. In fact, we find that the central charges of the $\widehat{\Gamma}(G)$ are given by
\begin{align}\label{eqn:centralcharges}
  \begin{aligned}
    a(\widehat{\Gamma}(G)) &= \frac{5}{24}\text{dim}(G) + \sum_i a(p_i, G) = \frac{p_4 - 1}{p_4} \text{dim}(G) - b(\widehat{\Gamma}(G)) - \frac{5}{48}f(\widehat{\Gamma}(G)) \,,\cr
    c(\widehat{\Gamma}(G)) &= \frac{1}{6}\text{dim}(G) + \sum_i c(p_i, G) = \frac{p_4 - 1}{p_4} \text{dim}(G) - \frac{1}{12}f(\widehat{\Gamma}(G)) \,.
  \end{aligned}
\end{align}

\begin{table}[H]
    \small
    \renewcommand{\arraystretch}{1.3}
    $ \mathbf{\widehat{D}_4}:\,
    \begin{array}{c|c|c|c|c|c|c}
        \multirow{2}{*}{$G$} & SU(N) & \multicolumn{2}{c|}{SO(N)} & \multirow{2}{*}{$E_6$} & \multirow{2}{*}{$E_7$} & \multirow{2}{*}{$E_8$} \\\cline{2-4}
         & N=2\ell & N=4\ell & N=4\ell+2 & & & \\\hline
        b(\widehat{D}_4(G)) & \frac{1}{8} & \frac{l}{4} & \frac{2\ell - 1}{8} & \frac{1}{4} & \frac{7}{8} & 1
    \end{array}$\\[6pt]
    $  \mathbf{\widehat{E}_6}:\
    \begin{array}{c|c|c|c|c|c}
        G & SU(3\ell) & SO(6\ell + 2) & E_6 & E_7 & E_8 \\\hline
        b(\widehat{E}_6(G)) & \frac{1}{6} & -\frac{1}{6} & \frac{1}{2} & \frac{1}{3} & \frac{2}{3}
    \end{array}$\\[6pt]
    $ \mathbf{\widehat{E}_7}:\
    \begin{array}{c|c|c|c|c|c|c}
        \multirow{2}{*}{$G$} & \multicolumn{2}{c|}{SU(N)} &  \multicolumn{4}{c}{SO(N)}\\\cline{2-7}
         & N=4\ell & N=4\ell-2 & N=8\ell & N=8\ell+2 & N=8\ell+4 & N=8\ell+6\\\hline
        b(\widehat{E}_7(G)) & \frac{3}{16} & \frac{1}{16} & \frac{l}{4} & \frac{4\ell-3}{16} & \frac{2\ell+1}{8} & \frac{4\ell+3}{16} \\\hline
        G &  E_6 & E_7 & E_8 \\\cline{1-4}
        b(\widehat{E}_7(G)) & -\frac{1}{8} & \frac{7}{16} & \frac{1}{2}
    \end{array}$\\[6pt]
    $  \mathbf{\widehat{E}_8}:\
    \begin{array}{c|c|c|c|c|c|c}
        \multirow{2}{*}{$G$} & \multicolumn{3}{c|}{SU(N)} & \multirow{2}{*}{$E_6$} & \multirow{2}{*}{$E_7$} & \multirow{2}{*}{$E_8$} \\\cline{2-4}
         & N=6\ell & N=6\ell+2,\ 6\ell+4 & N=6\ell+3 & & & \\\hline
        b(\widehat{E}_8(G)) & \frac{5}{24} & 
        \frac{1}{24} & 
        \frac{1}{12} & 
        \frac{1}{4} & 
        \frac{7}{24} & \frac{1}{3}\\
    \end{array}$\\
    $
    \begin{array}{c|c|c|c|c|c|c}
        \hline
        \multirow{2}{*}{\normalsize{$G$}} &  \multicolumn{6}{c}{SO(N)} \\\cline{2-7}
         & N=12\ell-4 & N=12\ell-2 & N=12\ell & N=12\ell+2 & N=12\ell+4 & N=12\ell+6 \\\hline
        b(\widehat{E}_8(G)) & \frac{3\ell - 1}{12} & \frac{2\ell - 1}{8} & \frac{\ell}{4} & \frac{6\ell - 5}{24} & \frac{3\ell+1}{12} & \frac{6\ell+1}{24}
    \end{array}$\\[2pt]
    \caption{The $\widehat{\Gamma}(G)$ theories have a subleading contribution to the central charge $a(\widehat{\Gamma}(G))$ which is encoded in the coefficient $b(\widehat{\Gamma}(G))$ as written in equation \eqref{eqn:centralcharges}. In these tables, we list the values of these coefficients, which are determined from the central charges of the gauged together $\mathcal{D}_p(G)$ theories.}
    \label{tbl:bofDE}
\end{table}

We note that the expressions in equation \eqref{eqn:centralcharges} simplify once both $b(\widehat{\Gamma}(G))$ and $f(\widehat{\Gamma}(G))$ vanish. 
In fact, these superfluous terms are non-zero only if at least one of the $p_i > 1$ divides the dual Coxeter number of $G$. The specific values that $b(\widehat{\Gamma}(G))$ takes, when it is non-zero, are written in the Table \ref{tbl:bofDE}.

Finally, we point out that for each of these four choices for $\widehat{\Gamma}$, the quantity $p_4$ is nothing other than the $\Delta_\Gamma$, as written in Table \ref{tbl:deltas}, associated to that element of the Deligne--Cvitanovi\'c exceptional series. In this way, we can see that the leading order contribution to the central charges has the form
\begin{equation}\label{eqn:acdeligne}
    a \sim c \sim \frac{\Delta_\Gamma - 1}{\Delta_\Gamma}\text{dim}(G) \,,
\end{equation}
as claimed in equation \eqref{eqn:lrgNc}.

\section{Infinitely many SCFTs with $a = c$}
\label{sec:aequalsc}

In this section, we explore the theories $\widehat{\Gamma}(G)$  which have $a = c$.\footnote{When a four-dimensional $\mathcal{N}=2$ SCFT has a Higgs branch that can be completely Higgsed, its quaternionic dimension is given as $d_\mathcal{H} = 24(c - a) = (n_h - n_v)$ where $n_h$ and $n_v$ are the (effective) number of hyper and vector multiplets. Theories with $a = c$ cannot have a Higgs branch where the Coulomb branch is completely Higgsed.} 
We first recall the formula for the central charges of $\widehat{\Gamma}(G)$ given in equation \eqref{eqn:centralcharges}. It is clear to see that a theory with $a = c$ occurs whenever we have
\begin{equation}\label{eqn:accond}
    f(\widehat{\Gamma}(G)) = 0 \qquad \text{and} \qquad b(\widehat{\Gamma}(G)) = 0 \,.
\end{equation}
The theories $\widehat{\Gamma}(G)$ satisfying $a=c$ are listed in Table \ref{tbl:allac} with their central charges. It is noteworthy that these central charges are all even integers and thus the central charges of the associated chiral algebras are a multiple of twenty-four:
\begin{equation}
    c_\text{2d} = -12c \in 24\mathbb{Z_-} \,.
\end{equation}
Aspects of some of the non-Lagrangian 4d $\mathcal{N}=2$ SCFTs in Table \ref{tbl:allac} have been explored in previous works. For example, the chiral algebra of the $\widehat{E}_6(SU(2))$ theory has been studied in \cite{Buican:2016arp,Buican:2020moo}; the construction of $\widehat{\Gamma}(G)$ from the minimal 6d $(1,0)$ SCFTs compactified on a torus has appeared in \cite{DelZotto:2015rca}. 

For a few specific choices of $\Gamma$ and $G$ the $\widehat{\Gamma}(G)$ theories have an alternative construction, as theories $J^b[k]$, as we have already discussed in Section \ref{sec:prop}. 
Among the theories with $a = c$ we study, we list the three examples where this `dual' description occurs in Table \ref{tbl:aclowrankdual}. Interestingly, a scan of the central charges of the $J^b[k]$ reveals that they generally have rational central charge $c$, instead of being an integer, and these three cases are the only examples we have found where there is overlap with $\widehat{\Gamma}(G)$.

\begin{table}[H]
\begin{threeparttable}
\centering
\renewcommand{\arraystretch}{1.4}
    $\begin{array}{c c}
    \toprule
      \phantom{SUNNN}\widehat{\Gamma}(G)\phantom{SUNNN} & \phantom{SUNNN}a=c\phantom{SUNNN} \\\midrule
      \widehat{D}_4(SU(2\ell + 1)) & 2\ell(\ell + 1) \\
      \widehat{E}_6(SU(3\ell \pm 1)) & 2\ell(3\ell \pm 2) \\
      \widehat{E}_6(SO(6\ell)) & 2\ell(6\ell + 1) \\
      \widehat{E}_6(SO(6\ell + 4)) & 2(2\ell + 1)(3\ell + 2) \\
      \widehat{E}_7(SU(4\ell \pm 1)) & 6\ell(2\ell \pm 1) \\
      \widehat{E}_8(SU(6\ell \pm 1))& 10\ell(3\ell \pm 1) \\\bottomrule
    \end{array}$
\end{threeparttable}
    \caption{All $\widehat{\Gamma}(G)$ theories satisfying $a=c$ with the values of their central charges, where $\ell$ is an arbitrary positive integer.}
    \label{tbl:allac}
\end{table}

\begin{table}[H]
    \begin{threeparttable}
    \centering
    \renewcommand{\arraystretch}{1.4}
    \begin{tabular}{cccc}
    \toprule
        $\widehat{\Gamma}(G)$ & $a = c$ & Coulomb branch operator dimensions & Alternative name \\\midrule
        $\widehat{E}_6(SU(2))$ & $2$ & $\left\{\frac{4}{3}\right\}^{\oplus 3} \oplus \{2\}$ & $D_4^6[3]=D_4^4[2]=(A_2, D_4)$ \\
        $\widehat{E}_7(SU(3))$ & $6$ & $\left\{\frac{5}{4}, \frac{9}{4} \right\}^{\oplus 2} \oplus \left\{ \frac{3}{2} \right\}^{\oplus 3} \oplus \{2, 3\}$ & $E_6^{12}[4] = (A_3, E_6)$ \\
        $\widehat{E}_8(SU(5))$ & $20$ & $\left\{\frac{3}{2}, \frac{5}{2}, \frac{4}{3}, \frac{7}{3}, \frac{10}{3}, \frac{5}{3} \right\}^{\oplus 2} \oplus \left\{\frac{7}{6}, \frac{13}{6}, \frac{19}{6}, \frac{25}{6}, 2, 3, 4, 5 \right\}$ & $E_8^{30}[6] = (A_5, E_8)$
        \\\bottomrule
    \end{tabular}
    \end{threeparttable}
    \caption{Physical properties and alternative constructions for some of the SCFTs $\widehat{\Gamma}(G)$ with $a = c$. These three theories are the only known theories which have an overlap with the $J^b[k]$ theories of \cite{Xie:2016evu}.}
    \label{tbl:aclowrankdual}
\end{table}

The theory with the lowest central charges in Table \ref{tbl:allac} is $\widehat{E}_6(SU(2))$, which has $a = c = 2$. This theory has a rank four Coulomb branch generated by operators of conformal dimensions
\begin{equation}
    \left\{\frac{4}{3}, \frac{4}{3}, \frac{4}{3}, 2 \right\} \,.
\end{equation}
In fact, this particular theory is rather well-known. It is composed via gauging together the diagonal subgroup of three copies of 
\begin{equation}
    \mathcal{D}_3(SU(2)) = H_1 \,,
\end{equation}
where on the right hand side we have the rank one Argyres--Douglas theory with $SU(2)$ flavor symmetry $H_1$. The $\widehat{E}_6(SU(2))$ theory is identical (or dual) to $D_4^6[3] =D_4^4[2]$ and $(A_2, D_4)$ in the notation of \cite{Wang:2015mra} and \cite{Cecotti:2010fi} respectively. 

Notice that the $\widehat{E}_6(SU(2))$ theory has several different realizations. Firstly, we obtain this theory via gauging three copies of the $\mathcal{D}_3(SU(2))=H_1$ theory. The $H_1=(A_1, A_3)=(A_1, D_3)$ theory has at least two 
$\mathcal{N}=1$ Lagrangian descriptions in terms of $SU(2)$ gauge theories resulting from the principal nilpotent Higgsing of $SU(4)$ \cite{Maruyoshi:2016aim} and a non-principal nilpotent Higgsing of $SO(8)$ \cite{Agarwal:2016pjo}. As a result, we can describe this $\widehat{E}_6(SU(2))$ theory as an $SU(2)^4$ gauge theory with appropriate matter contents and interactions. On the other hand, since this theory is identical (or dual) to $(A_2, D_4)$ theory, one can obtain another $\mathcal{N}=1$ Lagrangian description via a deformation of $SO(4)-USp(4)$ quiver gauge theory \cite{Agarwal:2017roi} 
\begin{align}\label{eqn:N1lag}
\begin{aligned}
\begin{tikzpicture}
\node[anchor=south west, draw, rectangle, inner sep=3pt, minimum size=5mm, text height=3mm](A0) at (0,0) {$SO(8)$};
\node[anchor=south west, draw,rounded rectangle, inner sep=3pt, minimum size=5mm, text height=3mm](A1) at (2.5,0) {$USp(4)$};
\node[anchor=south west, draw,rounded rectangle, inner sep=3pt, minimum size=5mm, text height=3mm](A3) at (5,0) {$SO(4)$};
\draw (A0)--(A1)--(A3);
\end{tikzpicture}
\end{aligned}\ .
\end{align}
This provides an IR duality of the $\widehat{E}_6(SU(2))$ theory. 

The $\widehat{E}_7(SU(3))$ theory has $a = c = 6$. This theory is obtained by gauging together two copies of $\mathcal{D}_4(SU(3))$ and one copy of $\mathcal{D}_2(SU(3)) = H_2$. Similar to the analysis for the $\widehat{E}_6(SU(2))$ theory, the $\widehat{E}_7(SU(3))$ theory is found to be (dual to) the $E_6^{12}[4]$ theory. We note that $\mathcal{D}_2(SU(3)) = H_2$ has a Lagrangian description, and in turn, we can write this $\widehat{E}_7(SU(3))$ theory with a partially-Lagrangian quiver. It would be interesting to see if there is a fully Lagrangian description to this theory. 

\subsection{The superconformal index}\label{sec:index}

In this section, we study the superconformal index \cite{Kinney:2005ej, Romelsberger:2005eg} of our $\widehat{\Gamma}(G)$ theories. The superconformal index captures the spectrum of short multiplets in a superconformal theory. 
We focus on $\Gamma = D_4, E_6, E_7, E_8$ in this section. 
Generally, the $\widehat{\Gamma}(G)$ theory involves non-Lagrangian $\mathcal{D}_p (G)$ Argyres--Douglas theory, which makes it rather difficult to evaluate the superconformal index in full generalities.   Apart from some special cases with known $\mathcal{N}=1$ Lagrangian descriptions \cite{Maruyoshi:2016tqk, Maruyoshi:2016aim, Agarwal:2016pjo}, the full indices for these theories are still lacking. 
Instead, we compute the Schur limit \cite{Gadde:2011uv,Gadde:2011ik} of the superconformal index, defined as
\begin{align}
    I_S (q) = \mathrm{Tr} (-1)^F q^{\Delta-R} \ , 
\end{align}
where $\Delta$ is the scaling dimension and $R$ is the Cartan of the $SU(2)_R$ symmetry. The Schur index for the $\mathcal{D}_p(G)$ theories having no extra flavor symmetry (i.e. $G$ is the only flavor symmetry of the theory) are computed in \cite{Song:2015wta, Xie:2016evu, Song:2017oew} to give
\begin{align} \label{eq:DpGidx}
    I_{\mathcal{D}_p(G)}(q, \vec{z}) =  \textrm{PE} \left[ \frac{q - q^p}{(1-q)(1-q^p)} \chi_{\textrm{adj}}^G (\vec{z}) \right] \ , 
\end{align}
where PE denotes the plethystic exponential, $\chi^G_{\textrm{adj}}$ is the character for the adjoint representation of the flavor symmetry $G$. When the flavor symmetry of the $\mathcal{D}_p(G)$ theory becomes larger than $G$, it has additional $U(1)$ factors.\footnote{It is possible to have a symmetry enhancement to a larger non-Abelian group as well.} In this case, we do not have a concise expression for the Schur index as in equation \eqref{eq:DpGidx}. Still, the Schur index for such theories can be computed using the method of topological quantum field theory  \cite{Buican:2015ina, Buican:2015tda, Song:2015wta, Buican:2017uka, Song:2017oew}.  In this section, we focus on the $\widehat{\Gamma}(G)$ theories without flavor symmetry that are built out of gauging $\mathcal{D}_p(G)$ theories without extra flavor symmetry. We further restrict ourselves to the theories with equal central charges $a=c$. 

We observe that the growth of the coefficients in the Schur index is rather slow, specifically they do not grow exponentially large. This is consistent with the formula for the Cardy-like limit \cite{DiPietro:2014bca,Buican:2015ina,Ardehali:2015bla,Cecotti:2015lab}
\begin{equation}
    \lim_{q \rightarrow 1} I(q) \sim e^{\frac{8\pi^2}{\omega}(c - a)} \underset{a=c}{\sim} 1 \,,
\end{equation}
where we have defined $\omega$ via $q = e^{-\omega}$. 
Even though the Schur limit of the index grows very slowly, the full superconformal index does exhibit exponential growth even for $a=c$ theories. 
The asymptotic expression for the `high-temperature' behavior of the \emph{full} superconformal index for 4d superconformal theory has been recently obtained to give \cite{Kim:2019yrz, Cabo-Bizet:2019osg, Cassani:2021fyv}
\begin{align}
    \lim_{\omega_{1, 2} \to 1} \textrm{log}I(\omega_1, \omega_2) \sim \frac{8 \Delta^3}{27 \omega_1 \omega_2} (5a-3c) + \frac{8\pi^2 \Delta}{3\omega_1 \omega_2} (a-c) \ , 
\end{align}
where $\Delta = \frac{\omega_1+\omega_2}{2} - \pi i$ and $\omega_{1, 2}$ are the chemical potentials for the combinations of angular momentum and $U(1)_R$ charge.\footnote{In order for this formula to hold, one actually needs to replace $(-1)^F$ by $(-1)^R$ in the definition of the index. It does not change the physical content of the index, which counts the short-multiplets up to recombination. However, it does simplify the expression to the above universal form.} In our case, we have $a=c \sim O(|G|)$ so that the \emph{full} superconformal index should exhibit exponential growth $e^{|G|}$.

\subsubsection*{$\widehat{D}_4(G)$ theories}

For the $\widehat{D}_4(G)$ theory, the Schur index can be computed using the following integral
\begin{align}\label{eqn:D4ind}
\begin{split}
        I_{\widehat{D}_4(G)}(q) &= \int [d\vec{z}]\ I_{\textrm{vec}}^G(q, \vec{z}) I_{\mathcal{D}_2(G)}(q, \vec{z})^4 \\
    &= \int [d\vec{z}]\ \textrm{PE} \left[ \left(\frac{-2q}{1-q} + \frac{4(q-q^2)}{(1-q)(1-q^2)} \right) \chi_{\textrm{adj}}^G (\vec{z})  \right] \\
     &= \int [d\vec{z}]\ \textrm{PE} \left[ \frac{2q - 2q^2}{1-q^2}\chi_{\textrm{adj}}^G (\vec{z})  \right] \ .
\end{split}
\end{align}
Here, we assumed that $p=2$ does not divide $h^\vee_G$ so that the $\mathcal{D}_2(G)$ does not have any other flavor symmetry besides $G$. 
A remarkable thing to notice here is that this Schur index is identical to that of the $\mathcal{N}=4$ super Yang--Mills theory with the gauge group $G$ (with flavor fugacity turned off) upon rescaling $q \to q^2$: 
\begin{align}
    I^{\mathcal{N}=4}_G (q) = \int [d\vec{z}]\, \textrm{PE}\left[ \frac{2q^{1/2} - 2q}{1-q} \chi_{\textrm{adj}}^G (\vec{z}) \right] .
\end{align}
This phenomenon is similar to the relation between the Schur index of $\mathcal{D}_2(G)$ theory versus that of a free hypermultiplet, where the former is given by the latter with $q\to q^2$ rescaling.

Let us list a few cases upon evaluating the integral explicitly:
\begin{subequations}\label{eq:D4index}
\begin{align} 
    I_{\widehat{D}_4(SU(3))} &= 1+3 q^2 +4 q^4 + 7 q^6+6 q^8+12 q^{10}+8 q^{12}+15 q^{14}+13 q^{16} + O(q^{22})\,, \\
    I_{\widehat{D}_4(SU(5))} &= 1+3 q^2+9 q^4+15 q^6+30 q^8+45 q^{10}+67 q^{12}+99 q^{14}+O\left(q^{16}\right)\,, \\
    I_{\widehat{D}_4(SU(7))} &= 1 + 3 q^2 + 9 q^4 + 22 q^6 + 42 q^8 + 81 q^{10} + 140 q^{12} + 231 q^{14} + O(q^{16})\,.
\end{align}
\end{subequations}
Remarkably, we find that the leading terms of the Schur indices are given by the generating function of MacMahon's generalized `sum-of-divisor' function \cite{macmahon1921divisors}, which is defined as
\begin{align}
    A_k (q) = \sum_{0 < m_1 <m_2 \cdots <m_k} \frac{q^{m_1 + \cdots m_k}}{(1-q^{m_1})^2 \cdots (1-q^{m_k})^2} \,.
\end{align}
For $k=1$, this becomes the generating function of the sum-of-divisor function
\begin{align}
    A_1 (q) = \sum_{n=1}^\infty \sigma_1(n)q^n = \sum_{m=1} \frac{q^m}{(1-q^m)^2} \ , 
\end{align}
where $\sigma$ is responsible for the name `sum-of-divisor':
\begin{align}
     \sigma_k (n) = \sum_{d|n} d^k \ . 
\end{align}
In terms of $A_k(q)$, which was shown to be quasi-modular in \cite{MR3028756}, we find that the Schur index for the $\widehat{D}_4 (SU(2k+1))$ theory is given as
\begin{align}
    I_{\widehat{D}_4 (SU(2k+1))}(q) = q^{-k(k+1)} A_k(q^2) \ . 
\end{align}
Notice that $a=c=2k(k+1)$ so that the prefactor is $q^{c_{2d}/24}$ where $c_{2d}=-12 c$ is the central charge for the associated chiral algebra or VOA. Also notice that the Schur index for $\mathcal{N}=4$ SYM theory with gauge group $SU(2k+1)$ is simply written in terms of $A_k(q)$, without rescaling:
\begin{align}
   I^{\mathcal{N}=4}_{SU(2k+1)}(q) = q^{-\frac{k(k+1)}{2}} A_k(q)   \,.
\end{align}
From this evidence of the quasi-modularity of the $\widehat{D}_4(SU(N))$ and $SU(N)$ $\mathcal{N}=4$ SYM (with $N$ odd), it is a natural extension to wonder if the other cases contain quasi-modularity (and hence number-theoretic properties for the Schur indices), which is yet to be explored.

We want to point out that the coefficients of $q^2$ in \eqref{eq:D4index} are all equally 3. One out of these three contributions to the $q^2$ term originates from the stress-tensor multiplet ($\widehat{\mathcal{C}}_{0(0, 0)}$ in Dolan--Osborn notation \cite{Dolan:2002zh}). We find the other two can only come from $\mathcal{D}_{\frac{1}{2}(0, \frac{1}{2})}$ and $\overline{\mathcal{D}}_{\frac{1}{2}(\frac{1}{2}, 0)}$ multiplets, which are fermionic. These three multiplets become generators when passed on to the associated VOA. We also notice that in the absence of a flavor current, the coefficient of the $q^2$ term can be either 1 or 3, since these three multiplets are the only possible short-multiplets in the Schur sector that can contribute to the $q^2$ term. The case of having 2 is excluded since it breaks parity invariance.

\subsubsection*{$\widehat{E}_n(G)$ theories}

The Schur index for the $\widehat{E}_6(G)$ theory is given as
\begin{align}\label{eqn:E6ind}
\begin{split}
    I_{\widehat{E}_6(G)} &= \int [d\vec{z}]\ I_{\textrm{vec}}^G(q, \vec{z}) I_{\mathcal{D}_3(G)}(q, \vec{z})^3 \\
        &= \int [d\vec{z}]\ \textrm{PE} \left[ \left(- \frac{2q}{1-q} +  \frac{3(q-q^3)}{(1-q)(1-q^3)} \right)\chi^G_{\textrm{adj}}(\vec{z}) \right] \\
         &= \int [d\vec{z}]\ \textrm{PE} \left[ \frac{q + q^2 - 2q^3}{1-q^3}  \chi^G_{\textrm{adj}}(\vec{z}) \right] \ , 
\end{split}
\end{align}
where we assume $p=3$ does not divide $h^\vee_G$. We list the first few terms for the indices of $\widehat{E}_6(G)$ theories of low rank:
\begin{subequations}
\begin{align}
I_{\widehat{E}_6(SU(2))} &= 1 + q^2 + q^3 + 2 q^6 + q^8 + q^{11} + 2 q^{12} + q^{15} + 2 q^{18} + 2 q^{20} +\cdots \,,\\
I_{\widehat{E}_6(SU(4))} &= 1 + q^2 + 2 q^3 + 2 q^4 + q^5 + 6 q^6 + 2 q^7 + 4 q^8 + 7 q^9 + 7 q^{10} + 4 q^{11} +\cdots \,,\\
I_{\widehat{E}_6(SU(5))} &= 1+q^2+2 q^3+2 q^4+2 q^5+7 q^6+2 q^7+8 q^8+10 q^9+8 q^{10}+ \cdots \,, \\ 
I_{\widehat{E}_6(SU(7))} &= 1+q^2+2 q^3+2 q^4+2 q^5+ 8 q^6+4 q^7+9 q^8+14 q^9+15 q^{10}+ \cdots \,, \\ 
I_{\widehat{E}_6(SO(8))} &= 1+q^2+q^3+2 q^4+2 q^5+4 q^6+3 q^7+6 q^8+5 q^9+8 q^{10}+7 q^{11}+\cdots \,.
\end{align}
\end{subequations}
Likewise, the Schur index for the $\widehat{E}_7(G)$ theory is given as
\begin{align}\label{eqn:E7ind}
    \begin{split}
    I_{\widehat{E}_7(G)} &= \int [d\vec{z}]\ I_{\textrm{vec}}^G(q, \vec{z}) I_{\mathcal{D}_2(G)}(q, \vec{z})I_{\mathcal{D}_4(G)}(q, \vec{z})^2 \\
        &= \int [d\vec{z}]\ \textrm{PE} \left[ \left(- \frac{2q}{1-q} +  \frac{q-q^2}{(1-q)(1-q^2)} + \frac{2(q-q^4)}{(1-q)(1-q^4)} \right)\chi^G_{\textrm{adj}}(\vec{z}) \right] \\
        &= \int [d\vec{z}]\ \textrm{PE} \left[ \frac{q + q^3 - 2q^4}{1-q^4}  \chi^G_{\textrm{adj}}(\vec{z}) \right] \ . 
\end{split}
\end{align}
As before, we assume $p=2, 4$ does not divide $h^\vee_G$. Upon evaluating the integral, we obtain the Schur indices. We list some cases where $G$ is a low rank $SU(N)$:
\begin{subequations}
\begin{align}
    I_{\widehat{E}_7(SU(3))} &= 1 + q^2 + q^3 + 2 q^4 + 3 q^6 + q^7 + 3 q^8 + q^9 + 3 q^{10} +\cdots \,,\\
    I_{\widehat{E}_7(SU(5))} &= 1+q^2+q^3+3 q^4+q^5+5 q^6+2 q^7+8 q^8+ \cdots \,, \\
    I_{\widehat{E}_7(SU(7))} &= 1+q^2+q^3+3 q^4+q^5+6 q^6+3 q^7+10 q^8+ 6 q^9 + 15 q^{10} + \cdots \,.
\end{align}
\end{subequations}
Finally, the Schur index for the $\widehat{E}_8(G)$ theory is given as
\begin{align}\label{eqn:E8ind}
    \begin{split}
    I_{\widehat{E}_8(G)} &= \int [d\vec{z}]\ I_{\textrm{vec}}^G(q, \vec{z}) I_{\mathcal{D}_2(G)}(q, \vec{z})I_{\mathcal{D}_3(G)}(q, \vec{z})I_{\mathcal{D}_6(G)}(q, \vec{z}) \\
        &= \int [d\vec{z}]\ \textrm{PE} \left[ \left(- \frac{2q}{1-q} + \sum_{p=2, 3, 6} \frac{q-q^p}{(1-q)(1-q^p)} \right)\chi^G_{\textrm{adj}}(\vec{z}) \right] \\
        &= \int [d\vec{z}]\ \textrm{PE} \left[ \frac{q + q^5 - 2q^6}{1-q^6}  \chi^G_{\textrm{adj}}(\vec{z}) \right] . 
\end{split}
\end{align}
For example, we get first terms of the Schur index for the $\widehat{E}_8(SU(5))$ and $\widehat{E}_8(SU(7))$ theories as
\begin{align}
I_{\widehat{E}_8(SU(5))} &= 1 + q^2 + q^3 + 2 q^4 + 2 q^5 + 4 q^6 + 2 q^7 + 6 q^8 +\cdots \ , \\ 
I_{\widehat{E}_8(SU(7))} &= 1 + q^2 + q^3 + 2 q^4 + 2 q^5 + 5 q^6 + 3 q^7 + 7 q^8 +\cdots \ .
\end{align}
Unfortunately, we do not find a closed-form expression for the indices of $\widehat{E}_n(G)$ theories except for the $\widehat{E}_6(SU(2))$ theory, which was already studied in \cite{Buican:2016arp} as a character for the $\mathcal{A}(6)$ chiral algebra of Feigin--Feigin--Tipunin \cite{Feigin:2007sp, MR2766985}. 

\subsection{A connection with the Schur index of $\mathcal{N}=4$ super Yang--Mills}\label{sec:N4index}

We observe an intriguing relationship between the Schur limit of the superconformal index of $\widehat{\Gamma}(G)$ and the Schur limit of the superconformal index of $\mathcal{N}=4$ super Yang--Mills with gauge group $G$. Writing this more concretely:
\begin{equation}\label{eqn:cool}
    I_{\widehat{\Gamma}(G)}(q) = I^{\mathcal{N}=4}_G(q^{\alpha_\Gamma}, q^{\alpha_\Gamma/2 - 1}) \,.
\end{equation}
Here the Schur index for $\mathcal{N}=4$ SYM with gauge group $G$ can be written as
\begin{equation}
    I^{\mathcal{N}=4}_G(q, x) = \int [\,d\vec{z}\,] \, \text{PE}\left[\left( -\frac{2q}{1 - q} + \frac{q^{\frac{1}{2}}}{1 - q}(x + x^{-1}) \right) \chi_\text{adj}^G(\vec{z}) \right] \,,
\end{equation}
where $[d\vec{z}]$ is the Haar measure for the group $G$ and $x$ is the fugacity of the additional $SU(2)$ flavor symmetry of $\mathcal{N}=4$ SYM from the $\mathcal{N}=2$ perspective. In order to compare with the $\widehat{\Gamma}(G)$ theories, we can rearrange to find that 
\begin{equation}\label{eqn:N4indspec}
    I^{\mathcal{N}=4}_G(q^{\alpha_\Gamma}, q^{\alpha_\Gamma/2 - 1}) = \int [\,d\vec{z}\,] \, \text{PE}\left[\left( \frac{q + q^{\alpha_\Gamma-1} - 2 q^{ \alpha_\Gamma}}{1 - q^{ \alpha_\Gamma}}  \right) \chi_\text{adj}^G(\vec{z}) \right] \,.
\end{equation}
We find that this relationship holds for all the $\widehat{\Gamma}(G)$ theories that lack any flavor symmetry, a superset of $a=c$ theories, which can be derived using Section \ref{sec:two} as the following:
\begin{equation}
  \begin{gathered}
    \widehat{\Gamma}(SU(N)) \qquad \text{with } \gcd(\alpha_\Gamma, N) = 1 \,, \cr
    \widehat{E}_6(SO(2N)) \,, \quad \widehat{D}_4(E_6)\,, \quad \widehat{E}_6(E_7)\,, \quad \widehat{E}_7(E_6) \,, \cr
    \widehat{D}_4(E_8)\,, \quad \widehat{E}_6(E_8)\,, \quad \widehat{E}_7(E_8) \,, \quad \widehat{E}_8(E_8) \,.
  \end{gathered}
\end{equation}
It is straightforward to see that $I_{\widehat{\Gamma}(G)}$ with $\Gamma=D_4,E_6,E_7,E_8$ given in equations \eqref{eqn:D4ind}, \eqref{eqn:E6ind}, \eqref{eqn:E7ind}, and \eqref{eqn:E8ind} indeed satisfy the relation \eqref{eqn:cool}.
A particular example of such a relationship has been studied for the theories $\widehat{\Gamma}(SU(N))$ with $\gcd(\alpha_\Gamma, N) = 1$ \cite{Buican:2020moo}.

It is perhaps surprising that the theories $\widehat{\Gamma}(G)$, which are obtained by gauging the diagonal $G$ of a collection of non-Lagrangian $\mathcal{D}_p(G)$ theories, have a Schur index which can be determined from the Schur index of $\mathcal{N}=4$ super Yang--Mills. We do not know any physical motivation for this correspondence, and we hope to return to this question in the future.
Inspired by \cite{Buican:2020moo}, we suspect that this identification of the Schur indices indicates the existence of a graded vector space isomorphism between the chiral algebras of the respective theories.

\section{Beyond $a=c$\,: Deligne exceptional series \& $G\in ADE$}\label{sec:ext1}

Thus far, we have established the 4d $\mathcal{N}=2$ SCFTs $\widehat{\Gamma}(G)$ that have $a = c$. For each $\widehat{\Gamma}$, the equality condition of the two central charges $a=c$ requires the constraints in equation \eqref{eqn:accond} to be satisfied, which in turn restrict the choices of $G$ to be those appearing in Table \ref{tbl:allac}. When we go beyond these choices of $G$, then we find theories that are no longer $a = c$.  

\subsection{$\Gamma = D_4, E_6, E_7, E_8$ with a generic $G$ of type $ADE$}

\begin{figure}[H]
\scriptsize
\centering
\begin{subfigure}[b]{0.8\textwidth}
\centering
\includegraphics{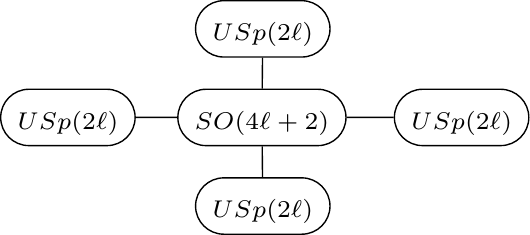}
\caption{The Lagrangian quiver description of  $\widehat{D}_4(SO(4\ell+2))$.}
\label{fig:d4quiverD}
\end{subfigure}\\[10pt]
\begin{subfigure}[b]{0.8\textwidth}
\centering
\includegraphics{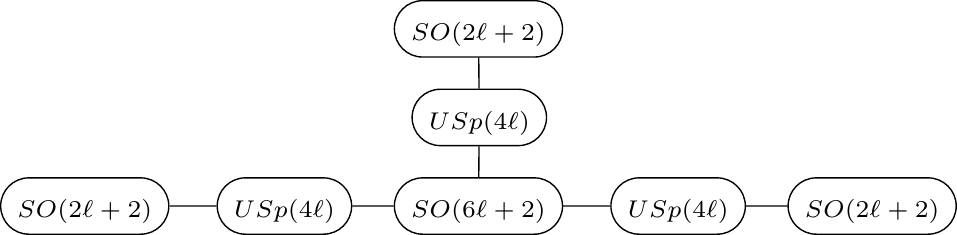}
\caption{The Lagrangian quiver description of  $\widehat{E}_6(SO(6\ell+2))$.}
\label{fig:e6quiverD}
\end{subfigure}\\[10pt]
\begin{subfigure}[b]{0.99\textwidth}
\centering
\includegraphics{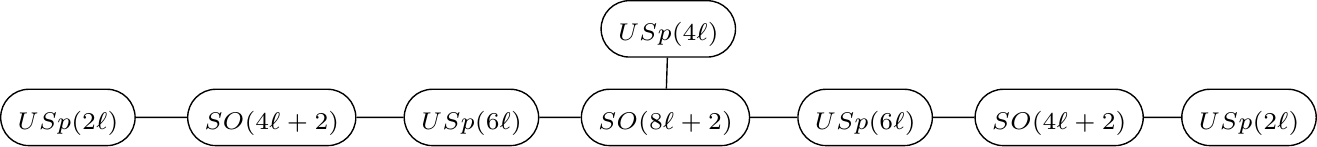}
\caption{The Lagrangian quiver description of  $\widehat{E}_7(SO(8\ell+2))$.}
\label{fig:e7quiverD}
\end{subfigure}\\[8pt]
\begin{subfigure}[b]{0.99\textwidth}
\centering
\includegraphics{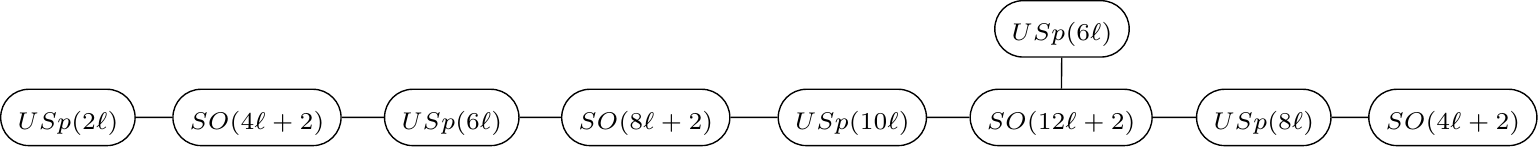}
\caption{The Lagrangian quiver description of  $\widehat{E}_8(SO(12\ell+2))$.}
\label{fig:e8quiverD}
\end{subfigure}
\caption{The SCFTs $\widehat{\Gamma}(SO(2\alpha_\Gamma\ell + 2))$ have Lagrangian quiver descriptions in terms of alternating $SO$ and $USp$ groups. We depict these quivers here.}
\label{fig:affquivDtype}
\end{figure}

For a theory $\widehat{\Gamma}(G)$ to have $a = c$, we have shown that a sufficient condition is that the dual Coxeter number ($h^\vee_G$) of $G$ and the largest comark ($\alpha_\Gamma$) of $\widehat{\Gamma}$ are co-prime. 
Then, to consider beyond these $a=c$ cases, we first examine the cases where the largest comark of $\widehat{\Gamma}$ divides the dual Coxeter number of $G$. These cases occasionally lead to Lagrangian quivers. If we take $G = SU(\alpha_\Gamma\ell)$ then we obtain the affine quiver gauge theories that we have seen in Figure \ref{fig:affquiv}. We also get Lagrangian quivers when we take 
\begin{align}
    G = SO(2\alpha_\Gamma\ell + 2),
\end{align}
which has dual Coxeter number $2\alpha_\Gamma\ell$; these quivers are shown in Figure \ref{fig:affquivDtype}. These quivers take the form of an affine $\widehat{\Gamma}$ Dynkin diagram where the gauge nodes are alternating $SO$ and $USp$ gauge groups.

We have considered the cases where $\gcd(h_G^\vee, \alpha_\Gamma) = \alpha_\Gamma$, where one typically obtains Lagrangian quivers, and at the other extreme we have considered the $\widehat{\Gamma}(G)$ theories which have $\gcd(h_G^\vee, \alpha_\Gamma) = 1$. The latter leads to theories which are non-Lagrangian and have $a = c$. Between these two extremes, we can consider the cases where 
\begin{align}
    1 < \gcd(h_G^\vee, \alpha_\Gamma) < \alpha_\Gamma.
    \label{eqn:gcdalphabound}
\end{align} 
These configurations can only exist when $\Gamma$ is $E_7$ or $E_8$. When $G = SU(N)$ and the condition in equation \eqref{eqn:gcdalphabound} is satisfied then one of the three legs of the $\widehat{\Gamma}(G)$ quiver becomes a Lagrangian quiver. The quiver description of these theories can be written as 
\begin{subequations}
\small
\begin{align}
\widehat{E}_7(SU(4\ell \pm 2)) \,&: \qquad 
\begin{aligned}
\begin{tikzpicture}
\node[anchor=south west, draw,rounded rectangle, inner sep=3pt, minimum size=5mm, text height=3mm](At) at (2,1) {$SU(2\ell \pm 1)$};
\node[anchor=south west](A2) at (-1.75,0) {$\mathcal{D}_4(SU(4\ell \pm 2))$};
\node[anchor=south west, draw,rounded rectangle, inner sep=3pt, minimum size=5mm, text height=3mm](A3) at (2,0) {$SU(4\ell \pm 2)$};
\node[anchor=south west](A4) at (4.5,0) {$\mathcal{D}_4(SU(4\ell \pm 2))$};
\draw (A2)--(A3)--(A4);
\draw (At)--(A3);
\end{tikzpicture}
\end{aligned}\\[0.5cm]
\widehat{E}_8(SU(6\ell \pm 2)) \,&: \qquad 
\begin{aligned}
\begin{tikzpicture}
\node[anchor=south west, draw,rounded rectangle, inner sep=3pt, minimum size=5mm, text height=3mm](At) at (2,1) {$SU(3\ell \pm 1)$};
\node[anchor=south west](A2) at (-1.75,0) {$\mathcal{D}_6(SU(6\ell \pm 2))$};
\node[anchor=south west, draw,rounded rectangle, inner sep=3pt, minimum size=5mm, text height=3mm](A3) at (2,0) {$SU(6\ell \pm 2)$};
\node[anchor=south west](A4) at (4.5,0) {$\mathcal{D}_3(SU(6\ell \pm 2))$};
\draw (A2)--(A3)--(A4);
\draw (At)--(A3);
\end{tikzpicture}
\end{aligned}\\[0.5cm]
\widehat{E}_8(SU(6\ell \pm 3)) \,&: \qquad 
\begin{aligned}
\begin{tikzpicture}
\node[anchor=south west](At) at (1.35,1) {$\mathcal{D}_2(SU(6\ell \pm 3))$};
\node[anchor=south west](A2) at (-1.75,0) {$\mathcal{D}_6(SU(6\ell \pm 3))$};
\node[anchor=south west, draw,rounded rectangle, inner sep=3pt, minimum size=5mm, text height=3mm](A3) at (2,0) {$SU(6\ell \pm 3)$};
\node[anchor=south west, draw,rounded rectangle, inner sep=3pt, minimum size=5mm, text height=3mm](A4) at (4.7,0) {$SU(4\ell \pm 2))$};
\node[anchor=south west, draw,rounded rectangle, inner sep=3pt, minimum size=5mm, text height=3mm](A5) at (7.55,0) {$SU(2\ell \pm 1))$};
\draw (A2)--(A3)--(A4)--(A5);
\draw (At)--(A3);
\end{tikzpicture}
\end{aligned}
\end{align}
\end{subequations}
A similar analysis can be performed for $G = SO(2N)$ and, depending on the value $\gcd(h_G^\vee, \alpha_\Gamma)$, some of the legs can become quivers with alternating $SO$ and $USp$ gauge nodes, similar to Figure \ref{fig:affquivDtype}.

A subset of the theories with $a \neq c$ that we are interested in are those that lack any flavor symmetry. For such theories, it is particularly simple to determine the Schur index, as we have seen in Section \ref{sec:N4index}. Here we briefly focus on the class of examples with $G = E_8$, as $\widehat{\Gamma}(E_8)$ theories never have any flavor symmetry, regardless of the choice of $\Gamma$. First, we note that the difference of the central charges for each theory is given by
\begin{equation}
    \begin{array}{c|c|c|c|c}
        \Gamma & D_4 & E_6 & E_7 & E_8 \\\hline
        24(c - a) & 24 & 16 & 12 & -8 
    \end{array}\ .
\end{equation}
For $\widehat{E}_8(E_8)$ we notice that $a > c$, indicating the presence of fermionic generators of the associated VOA or chiral algebra \cite{Buican:2016arp}.  Computing the explicit expansion of the Schur indices for each of these four theories is straightforward, in principle, using the methods explained in Section \ref{sec:index}. However, they are computationally intensive and we leave their determination for future work.

\subsection{Deligne--Cvitanovi\'c exceptional series}

The form of the central charges obtained for the theories $\widehat{\Gamma}(G)$ with $\Gamma = D_4, E_6, E_7$, and $E_8$ in Section \ref{sec:prop} is highly suggestive. The central charge $c$ for the theory $\widehat{\Gamma}(G)$ behaves like 
\begin{equation}\label{eqn:swiper}
    c \sim \frac{\Delta_\Gamma - 1}{\Delta_\Gamma}\text{dim}(G) \,,
\end{equation}
where $\Delta_\Gamma$ are the rational numbers (written in Table \ref{tbl:deltas}) associated to the groups $\Gamma$ appearing in the Deligne--Cvitanovi\'c exceptional series. A natural question to wonder is then if there exist superconformal gauge theories, with a single gauge group $G$ and coupled to some Lagrangian or non-Lagrangian matter, where the central charge $c$ has the form of equation \eqref{eqn:swiper} for the other simply-laced $\Gamma$ in the Deligne--Cvitanovi\'c exceptional series. We find that such theories do exist. 

If there exists a Lagrangian description of such a quiver gauge theory, then the central charges, $a$ and $c$, are given in terms of the number of vector multiplets $n_v$ and the number of (full) hypermultiplets $n_h$ in that Lagrangian description:
\begin{align}
    \ a = \frac{5}{24} n_v + \frac{1}{24} n_h,\quad c= \frac{1}{6} n_v + \frac{1}{12} n_h .
    \label{eqn:TALLJAEWON}
\end{align}
For non-Lagrangian theories, $n_v$ and $n_h$ can be non-integers and can be regarded as the `effective' numbers of vectors and hypers. Since we require the theories to have a $G$ gauge sector, we must have at least $n_v \geq \text{dim}(G)$.

The putative $\widehat{H}_0(G)$ SCFT should have a central charge of the form
\begin{equation}
    c \sim \frac{1}{6} \text{dim}(G) \,.
\end{equation}
We can see from the equation \eqref{eqn:TALLJAEWON} that this $c$ is precisely the central charge of a free vector multiplet. In four-dimensions, this constitutes a (free) SCFT, and thus we determine this to be the theory corresponding to the $\widehat{H}_0(G)$ theory:
\begin{equation}
    \widehat{H}_0(G) \, : \qquad \mymk{G} \ .
\end{equation}

Next, we determine a candidate for the $\widehat{H}_1(G)$ SCFT. The central charge should have the following form:
\begin{equation}
    c \sim \frac{1}{4}\text{dim}(G) = \frac{1}{6}\text{dim}(G) + \frac{1}{12}\text{dim}(G) \,.
\end{equation}
The latter equality, together with equation \eqref{eqn:TALLJAEWON}, makes it clear that such a central charge can also be engineered with a Lagrangian theory -- a theory of a single $G$-vector multiplet and a massless adjoint hypermultiplet. This matter spectrum satisfies the vanishing condition of the one-loop beta function. In fact, this theory also has an enhanced supersymmetry; it is the well-known description of $\mathcal{N}=4$ super Yang--Mills. Henceforth we define the $\widehat{H}_1(G)$ theory to be
\begin{equation}
    \widehat{H}_1(G) \, : \qquad \vcenter{\hbox{\includegraphics[scale=0.3]{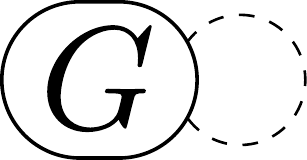}}} \, ,
\end{equation}
which does satisfy the desired feature of the scaling of the central charge $c$ in equation \eqref{eqn:swiper}. 

For the putative $\widehat{H}_2(G)$ theories, we are able to achieve our goal only for the classical $G$. We search for theories which have
\begin{equation}
    c \sim \frac{1}{3}\text{dim}(G) = \frac{1}{6}\text{dim}(G) + \frac{1}{6}\text{dim}(G) \,,
\end{equation}
which implies that we should have
\begin{equation}\label{eqn:h2sim}
    n_h \sim 2 \text{dim}(G) \, ,
\end{equation}
if we demand a Lagrangian description. 
Having two adjoint hypermultiplets seems like a good choice, however the one-loop beta function does not vanish with such a matter spectrum. Instead, we find that $G = SU(K)$ gauge theory with $2K$ fundamental hypermultiplets, which does have vanishing beta function, leads to 
\begin{equation}
    n_h = 2 K^2 = 2(K^2 - 1) + 2 = 2\, \text{dim}(SU(K)) + 2 \,.
\end{equation}
This produces the correct coefficient in the central charge $c$ at leading order in $K$. A similar analysis can be done for $G = SO(2K)$ with $4K-4$ half-hypermultiplets in the vector representation. We then have
\begin{equation}
    n_h = \frac{(2K)(4K-4)}{2} = 2 \left(\frac{2K(2K-1)}{2} \right) - 2K = 2\, \text{dim}(SO(2K)) - 2K \,.
\end{equation}
Again, this produces the desired central charge at leading order, consistent with equation \eqref{eqn:h2sim}. This construction does not extend to theories with exceptional $G$, since we lack the number of hypermultiplets to fulfill the conformality condition. To illustrate this point, let us provide an extreme example when $G=E_8$. For the $G=E_8$ theory, the number of hypermultiplets is $n_h=\textrm{dim}(G)$, which is smaller than the required amount in equation \eqref{eqn:h2sim}. Moreover, the large $K$ limit is not available. It would be interesting if there exists a theory with exceptional gauge group that achieves equation \eqref{eqn:h2sim} from gauging a (partially) non-Lagrangian theory.  


\section{Beyond $a=c$\,: generic $G$ and $\Gamma$}\label{sec:gDM}

We have established thus far a collection of 4d $\mathcal{N}=2$ SCFTs that we denote as $\widehat{\Gamma}(G)$. As it is illustrated in Figure \ref{fig:affquiv}, the theories
\begin{equation}
    \widehat{D}_4(SU(2\ell))\,,\quad \widehat{E}_6(SU(3\ell))\,,\quad \widehat{E}_7(SU(4\ell))\,,\quad  \widehat{E}_8(SU(6\ell))\,,
    \label{eqn:Lagrangianaequalc}
\end{equation}
correspond to Lagrangian quivers. In particular, these are the cases where the gauge group is  simply given by
\begin{align}
    G = SU(\alpha_\Gamma \ell) \,,
\end{align}
where $\ell$ is an arbitrary positive integer and $\alpha_\Gamma$ is the largest comark of $\widehat{\Gamma}$. These affine Dynkin diagram shaped quivers can be realized in string theory as the worldvolume theory on a stack of $\ell$ D3-branes probing a $\mathbb{C}^2/\Gamma$ singularity \cite{Douglas:1996sw}.  
We refer to these quivers as affine quiver gauge theories, or just affine quivers associated to $\widehat{\Gamma}$. 

\begin{figure}[H]
\centering
\begin{subfigure}[b]{0.49\textwidth}
\centering
\includegraphics{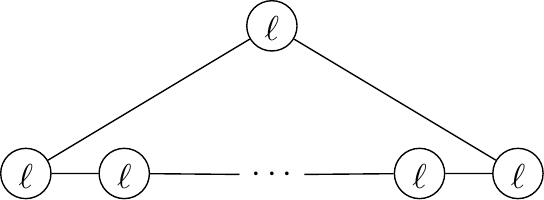}
\caption{The $\widehat{A}_{N-1}(SU(\ell))$ quiver.}
\label{fig:ADM}
\end{subfigure}\hspace{2mm}
\begin{subfigure}[b]{0.49\textwidth}
\centering
\includegraphics{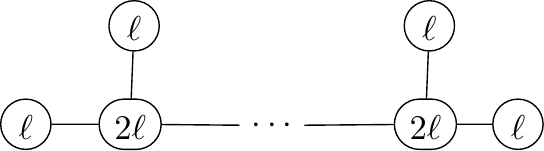}
\caption{The $\widehat{D}_{N+4}(SU(2\ell))$ quiver.}
\label{fig:DDM}
\end{subfigure}\vspace{4mm}
\begin{subfigure}[b]{0.49\textwidth}
\centering
\includegraphics{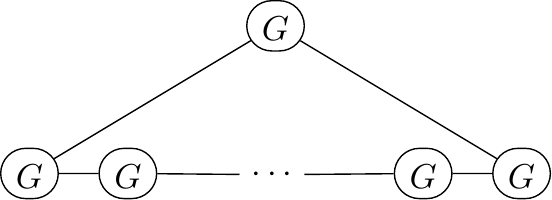}
\caption{The generalization of the $\widehat{A}_{N-1}(SU(\ell))$ quiver to $\widehat{A}_{N-1}(G)$.}
\label{fig:ADMg}
\end{subfigure}\hspace{2mm}
\begin{subfigure}[b]{0.49\textwidth}
\centering
\includegraphics{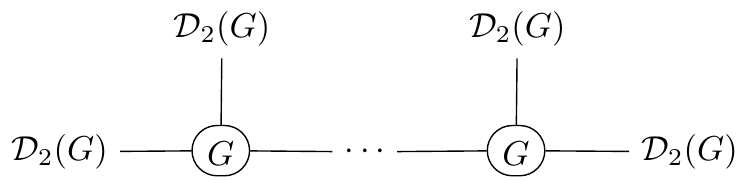}
\caption{The generalization of the $\widehat{D}_{N+4}(SU(2\ell))$ quiver to $\widehat{D}_{N+4}(G)$.}
\label{fig:DDMg}
\end{subfigure}
\caption{We show the affine quivers associated to the affine Dynkin diagrams $\widehat{A}_{N-1}$ and $\widehat{D}_{N+4}$, together with the generalizations that we consider. We can take $N \geq 1$ and $N \geq 0$ respectively, and we remind the reader that an integer $K$ inside a gauge node implies the gauge group $SU(K)$.}
\label{fig:DMgen}
\end{figure}

The affine quivers exist for all affine ADE Dynkin diagrams, not just for $D_4$, $E_6$, $E_7$, and $E_8$. Then a natural generalization to the $\widehat{\Gamma}(G)$ construction is to allow $\Gamma$ to be any of type ADE. In this section, we introduce a generalization of the remaining affine quivers to non-Lagrangian theories: we consider $\widehat{\Gamma}$ an arbitrary affine ADE Dynkin diagram and $G$ an arbitrary ADE Lie group. In Figures \ref{fig:ADM} and \ref{fig:DDM}, we depict the Lagrangian affine quivers for $\widehat{A}_N$ and $\widehat{D}_{N+4}$, respectively. Based on these, we draw the generalization to arbitrary $\Gamma$ and $G$ that we shall consider in Figures \ref{fig:ADMg} and \ref{fig:DDMg}.

The first subtlety that arises in the generalization, that we can see directly in Figures \ref{fig:ADMg} and \ref{fig:DDMg}, is that we have gauge nodes connected together as $\mymk{G} - \mymk{G}$. When $G$ is $SU(K)$, the standard formulation of 4d $\mathcal{N}=2$ superconformal quiver gauge theory is to include a bifundamental hypermultiplet for each link between two gauge nodes. Since we consider an arbitrary group $G$ of type ADE, we need to clarify the notion that corresponds to the links between two gauge nodes $G$. Extending the concept of bifundamental matter, each link corresponds to the $G\times G$ theory \cite{Ohmori:2015pua}. It is easy to check that when $G=SU(K)$ this generalization does indeed give the bifundamental hypermultiplet.

The $G \times G$ generalized bifundamentals can be constructed from both 6d $(2,0)$ SCFTs on a punctured Riemann surface and 6d $(1,0)$ SCFTs on a torus.\footnote{In fact, the existence of both 6d $(2,0)$ and 6d $(1,0)$ origins holds not just for the $G \times G$ generalized bifundamental, but also for all nilpotent Higgsing of $G \times G$, as was shown in \cite{Baume:2021qho}.} From the class $\mathcal{S}$ perspective, it can be realized as the 6d $(2,0)$ SCFT of type $G$ compactified on a sphere with two full punctures and one simple puncture \cite{Chacaltana:2010ks,Chacaltana:2012zy,Chacaltana:2014jba,Chacaltana:2017boe,Distler:2017xba,Distler:2018gbc}. The $G \times G$ flavor symmetry arises from the two full punctures (F), which correspond to the maximal nilpotent orbits of $G$, and the simple puncture corresponding to the subregular nilpotent orbit does not contribute to the flavor symmetry. The flavor central charge of each $G$ is $k_G = 2h_G^\vee$, and thus we can conformally gauge two such $G$s together. 

We note that this $G\times G$ theory can alternatively be constructed from the 6d $(1,0)$ SCFTs known as rank-one $(G, G)$ conformal matter \cite{DelZotto:2014hpa}. Via torus-compactification, this yields a minimal conformal matter of type $(G,G)$ in the resulting 4d $\mathcal{N}=2$ theory \cite{Ohmori:2015pua}.
This construction can be realized with a string-theoretic origin as the worldvolume theory of an M5-brane probing an ADE orbifold, compactified on a torus. 

Using the generalized notion of bifundamentals, we can now define the generalized affine quiver theories, that appear in Figures \ref{fig:ADMg} and \ref{fig:DDMg}. For these generalized theories, we are interested in computing the central charges $a$ and $c$, that are no longer identical to each other. To this end, we need to determine the central charges of the $G \times G$ generalized bifundamental theories, which are computed from either the class $\mathcal{S}$ or the 6d $(1,0)$ on $T^2$ perspectives. They are found to be
\begin{equation}
    \begin{aligned}
        a^{G} &= \frac{1}{24} \left(6\Lambda_G(r_G + 1) - 5(d_G + 1)\right) \,, \cr
        c^{G} &= \frac{1}{12} \left(3\Lambda_G(r_G + 1) - 2(d_G + 1)\right) \,,
    \end{aligned}
\end{equation}
where $d_G$ is the dimension of $G$, $r_G$ is the rank of $G$, and $\Lambda_G$ denotes the order of the finite ADE subgroup of $SU(2)$ associated to $G$ \cite{Ohmori:2014kda, Intriligator:2014eaa}. These quantities are summarized in Table \ref{tbl:gpdata}. As expected, when $G = SU(K)$, the central charges become
\begin{align}
    (24a^{SU(K)}, 12c^{SU(K)}) = (K^2, K^2),
\end{align}
which are exactly the central charges of a free bifundamental hypermultiplet.

\subsection{$\widehat{A}_{N-1}$-series}

The $\widehat{A}_{N-1}(G)$ quivers are circular quivers containing $N$ gauge nodes associated with the gauge group $G$ and $N$ links between pairs of gauge nodes that are associated to the $G \times G$ generalized bifundamentals. With this structure, it is straightforward to determine the central charges $a$ and $c$ for the $\widehat{A}_{N-1}(G)$ theories as
\begin{equation}
    \begin{aligned}
        a(\widehat{A}_{N-1}(G)) &= \frac{5N}{24}d_G + N a^G = \frac{N}{24} \left(6\Lambda_G (r_G + 1) - 5\right) \,, \cr
        c(\widehat{A}_{N-1}(G)) &= \frac{N}{6}d_G + N c^G = \frac{N}{12} \left(3\Lambda_G (r_G + 1) - 2\right) \,.
    \end{aligned}
\end{equation}
For these theories, we find that the difference of the central charges does not depend on the choice of $G$; it is simply
\begin{equation}
    c(\widehat{A}_{N-1}(G)) - a(\widehat{A}_{N-1}(G)) = \frac{N}{24} \,.
\end{equation}
We want to emphasize that this is strictly positive and thus $a\neq c$ for all the $\widehat{A}_{N-1}(G)$ theories.

Except for the case where $G = SU(K)$, the theories $\widehat{A}_{N-1}(G)$ lack any flavor symmetry. As the $\widehat{A}_{N-1}(G)$ theories are obtained by gauging together 4d $(G, G)$ conformal matter theories pairwise, there is a class $\mathcal{S}$ construction for $\widehat{A}_{N-1}(G)$.  Each of the conformal matter theories is obtained by compactifying the 6d $(2,0)$ SCFT of type $G$ on a sphere with two full punctures and a simple puncture, and the gauging procedure involves joining two of the full punctures with a cylinder. In the end, the circular quiver $\widehat{A}_{N-1}(G)$ can be constructed as the compactification of the 6d $(2,0)$ SCFT of type $G$ on a torus with $N$ simple punctures. Other than $G=SU(K)$, they are all non-Lagrangian theories for $N>1$. When $N=1$, it is identical to $\mathcal{N}=4$ SYM theory of gauge group $G$ with an additional free hypermultiplet \cite{Gaiotto:2012uq}.
Furthermore, the class $\mathcal{S}$ realization allows us to compute the (Schur limit of the) superconformal index utilizing the 4d/2d correspondence of \cite{Gadde:2011ik, Gadde:2011uv}.

\subsection{$\widehat{D}_{N+4}$-series}

Finally, let us direct our attention to the $\widehat{D}_{N+4}(G)$ generalized affine quivers. These are built from four $\mathcal{D}_2(G)$ theories, $N + 1$ gauge nodes, and $N$ copies of the $G \times G$ generalized bifundamental. Putting these constituents together one can see that the central charges are
\begin{equation}
    \begin{aligned}
        a(\widehat{D}_{N+4}(G)) &= a(\widehat{D}_4(G)) + \frac{5N}{24}d_G + N a^G = a(\widehat{D}_4(G)) + \frac{N}{24} \left(6 \Lambda_G (r_G + 1) - 5\right) \,,\cr
        c(\widehat{D}_{N+4}(G)) &= c(\widehat{D}_4(G)) + \frac{N}{6}d_G + N c^G = c(\widehat{D}_4(G)) + \frac{N}{12} \left(3 \Lambda_G (r_G + 1) - 2\right) \,.
    \end{aligned}
\end{equation}
We note that we write the central charges here in terms of the central charges of the $\widehat{D}_4(G)$ theories determined in equation \eqref{eqn:centralcharges}.
Other than the special case of $\widehat{D}_4(SU(2\ell + 1))$, the theories $\widehat{D}_{N+4}(G)$ for all choices of $G$ with $N>0$ have some flavor symmetry. 
The Schur index for the $\widehat{D}_{N+4}(SU(K))$ theory can be easily determined from the known result for the theory $\mathcal{D}_2(SU(2\ell+1))$, given as equation \eqref{eq:DpGidx}, and from the fact that the $\mathcal{D}_2(SU(2\ell))$ theory is simply identical to the Lagrangian $SU(\ell)$ gauge theory with $2\ell$ fundamental hypermultiplets. 

For other choices of $G$, the Schur index is not known in full generality. One of the methods for computing the index appeals to the TQFT structure of the index, coming from the class $\mathcal{S}$ realization of $\mathcal{D}_p (G)$ theories. For the current case, we need to know the wave function for the irregular puncture with flavors, which is unavailable except for $G=SU(K)$ case \cite{Buican:2017uka}.   
We leave the computation of the indices for these theories as a future work. 

Another thing to note is that the $\widehat{D}_{N+4}(G)$ and $\widehat{E}_N(G)$ SCFTs do not have known class $\mathcal{S}$ constructions, unlike the $\widehat{A}_{N-1}(G)$ SCFTs, whose class $\mathcal{S}$ construction is simply given by the $N$-punctured torus. It would be interesting to look for a six-dimensional origin of the other $\widehat{\Gamma}(G)$ theories.  

\subsection{An alternative link: non-minimal bifundamentals}

As a final remark in this section, we provide another possible option for the kind of the theory that we can associate to each link. As we discussed at the opening of Section \ref{sec:gDM}, when generalizing the affine quivers from affine Dynkin diagrams involving $SU(K)$ gauge nodes to those involving gauge nodes $G$, we are required to specify what we mean by the link between two gauge nodes. As we discussed earlier in this section, a natural generalization to an $SU(K)^2$ bifundamental hypermultiplet is the 4d $\mathcal{N} = 2$ SCFTs known as $(G, G)$ conformal matter. These theories do exist for all ADE groups $G$, and in a certain sense they are the ``minimal'' theories that one can introduce as links. However, when $G = SO(2K)$, there is another option for the link between two gauge nodes. This is to include a Lagrangian theory of the form
\begin{align}
\begin{tikzpicture}[baseline={([yshift=-.5ex]current bounding box.center)}]
\node[anchor=south west, draw,rounded rectangle, inner sep=3pt, minimum size=5mm, text height=3mm](A1) at (0,0) {$SO(2K)$};
\node[anchor=south west, draw,rounded rectangle, inner sep=3pt, minimum size=5mm, text height=3mm](A5) at (2.3,0) {$SO(2K)$};
\draw (A1)--(A5);
\end{tikzpicture}
\quad = \quad
\begin{tikzpicture}[baseline={([yshift=-.5ex]current bounding box.center)}]
\node[anchor=south west, draw,rounded rectangle, inner sep=3pt, minimum size=5mm, text height=3mm](A1) at (0,0) {$SO(2K)$};
\node[anchor=south west, draw,rounded rectangle, inner sep=3pt, minimum size=5mm, text height=3mm](A3) at (2.4,0) {$USp(2K-2)$};
\node[anchor=south west, draw,rounded rectangle, inner sep=3pt, minimum size=5mm, text height=3mm](A5) at (5.75,0) {$SO(2K)$};
\draw[dashed] (A1)--(A3)--(A5);
\end{tikzpicture}\ .
\end{align}
In this setting, each of the dashed links on the right-hand-side is simply a bifundamental half-hypermultiplet. We shall refer to these kind of links as \textit{orthosymplectic} (OSp) links. 

The central charges of these orthosymplectic links can be determined straightforwardly from their Lagrangian descriptions and utilizing equation \eqref{eqn:TALLJAEWON}:
\begin{empheq}[left=\empheqlbrace]{align}
\begin{aligned}
    \ a^\text{OSp(2K)} &= \frac{1}{24}(K-1)(14K-5),\\
    c^\text{OSp(2K)} &= \frac{1}{6}(K-1)(4K-1).
\end{aligned}
\end{empheq}
It is interesting to note that the difference between the central charge $a$ of the orthosymplectic link and that of the conformal matter link is
\begin{equation}
    a^\text{OSp(2K)} - a^{SO(2K)} = \frac{29}{12} > 0 \,,
\end{equation}
which can be taken as an evidence that the conformal matter link is a ``more minimal'' link. Another piece of evidence is that in the class $\mathcal{S}$ perspective, the orthosymplectic link requires an additional puncture, namely the twisted null puncture \cite{Tachikawa:2009rb}. Therefore, the $\widehat{A}_{N-1}$-type quivers with orthosymplectic links are realized from a torus with $N$ (untwisted) minimal punctures, and $N$ twisted null punctures. 

For a circular quiver with $N$ gauge nodes and $G=SO(2K)$, as depicted in Figure \ref{fig:ADMg}, connected by this kind of \textit{orthosymplectic matter}, we find that the central charges are
\begin{empheq}[left=\empheqlbrace]{align}
\begin{aligned}
    \ a &= \left(K(K-1) + \frac{5}{24}\right) N,\\
    c &= \left(K(K-1) + \frac{1}{6}\right) N .
\end{aligned}
\end{empheq}
The difference between these central charges is
\begin{align}
    c-a=-\frac{N}{24} \,,
\end{align}
which is strictly negative and its absolute value can be arbitrarily large by increasing the number of $SO(2K)$ gauge nodes in the circular quiver. We note that these theories have $a > c$, which implies existence of fermionic generators in the associated VOA \cite{Buican:2016arp}. A similar determination of the central charges can be made for the $\widehat{D}$-type quivers, shown in Figure \ref{fig:DDMg}, where the links are taken to be the orthosymplectic matter instead of the conformal matter. 

\section{Discussion}
\label{sec:discussion}

We have constructed a set of four-dimensional $\mathcal{N}=2$ superconformal theories $\widehat{\Gamma}(G)$, labeled by a pair of ADE groups $\Gamma$ and $G$. For a generic choice of $\Gamma$ and $G$, these theories involve Argyres--Douglas and  conformal matter theories, and thereby do not admit weakly-coupled Lagrangian descriptions. 

Among them, theories with $\Gamma = D_4, E_6, E_7, E_8$ exhibit particularly interesting aspect. One of the fascinating features of these (strictly $\mathcal{N}=2$) SCFTs is their similarity to $\mathcal{N}=4$ super Yang--Mills. When $\gcd(\alpha_\Gamma, h_G^\vee) = 1$ with $\alpha_\Gamma$ being the largest comark of the affine Lie group $\Gamma$, then this similarity is manifest in the central charges $a$ and $c$:
\begin{equation}
    a(\widehat{\Gamma}(G)) = c(\widehat{\Gamma}(G)) \sim d_G \,.
\end{equation}
We emphasize that there has been almost no known examples of genuinely $\mathcal{N}=2$ SCFTs with equal central charges $a=c$. 
For a holographic theory, the difference between two central charges ($a-c$) tends to be subleading in the $1/N$ expansion, but it does give a non-trivial correction to the bulk action. For instance, this can lead to the violation of the celebrated entropy-viscosity ratio bound \cite{Kovtun:2004de, Buchel:2008vz}. 
Moreover, for the theories $\widehat{\Gamma}(G)$ with no flavor symmetry, the Schur index of such a $\widehat{\Gamma}(G)$ is identical to that of the $\mathcal{N}=4$ SYM theory upon rescaling of parameters! More precisely, we find that
\begin{align}
    I_{\widehat{\Gamma}(G)}(q) = I^{\mathcal{N}=4}_G(q^{\alpha_\Gamma}, q^{\alpha_\Gamma/2 - 1}) \,.
\end{align}
It would be interesting to explore the implications of this correspondence to determine more about the $\widehat{\Gamma}(G)$ theories, such as their associated VOAs \cite{Buican:2020moo} or their holographic duals.

In fact, when $G=SU(\alpha_\Gamma\ell)$, we notice that the theory $\widehat{\Gamma}(SU(\alpha_\Gamma\ell))$ is the well-studied affine quiver gauge theory describing the worldvolume theory on $\ell$ D3-branes probing $\mathbf{C}^2/\Gamma$ singularity, which is depicted in Figure \ref{fig:affquiv}. Its holographic dual is well-understood from the pioneering example of the AdS/CFT correspondence as
\begin{equation}
    \text{AdS}_5 \times S^5/\Gamma
\end{equation}
with $\ell$ units of five-form flux through the $S^5/\Gamma$ \cite{Kachru:1998ys}. We note that the quotient is the finite ADE group $\Gamma$. Then it is natural to expect a simple holographic dual description (or a type IIB realization) of our $\widehat{\Gamma}(SU(N))$ theories. When $N$ is not divisible by $\alpha_\Gamma$ for $\Gamma = D_4, E_6, E_7, E_8$, one of the necessary aspects for such a theory $\widehat{\Gamma}(SU(N))$ is that the singularities associated to $\Gamma$ have to be ``frozen'' so that one cannot have any exactly marginal operators besides the coupling for the $SU(N)$ that we gauge. When considering affine quivers, all the gauge couplings are marginal and it follows that there are $r_\Gamma + 1$ exactly marginal couplings. It also looks like a ``fractionalization'' of $N$ into $N=\alpha_\Gamma \ell + m$, so that we have $\ell$ number of D3-branes with extra $m/\alpha_\Gamma$. It would be extremely interesting to find a holographic dual or a string-theoretic realization for the $\widehat{\Gamma}(G)$ theories for generic choices of $\Gamma$ and $G$.

The $\mathcal{D}_p(G)$ Argyres--Douglas theories considered in this paper have a 6d realization given by $(2, 0)$ theory of type $G$ compactified on a sphere with a regular puncture and an irregular puncture.\footnote{We note that the same Argyres--Douglas theory may have several different class $\mathcal{S}$ realizations.} In particular, the $\widehat{A}_{N-1}(G)$ theory can be realized from 6d $(2, 0)$ theory of type $G \in ADE$ on a torus with $N$ punctures. However, we are not aware of a complete geometric picture of the $\widehat{\Gamma}(G)$ theories other than $\Gamma = A_{N-1}$, which requires gluing (gauging) $\mathcal{D}_p(G)$ theories. 

On the other hand, some of the theories we discussed in the paper have their realizations in terms of 6d $(1, 0)$ SCFTs. For instance, the aforementioned  $\widehat{A}_{N-1}(G)$ theory that has 6d $(2,0)$ origin can also be obtained from 6d $(1, 0)$ theory formed out of gluing $N$ copies of conformal matter and reducing on a torus. The 6d rank $N$ conformal matter theory of type $(G,G)$, where $G$ is an ADE Lie group, is the theory that lives on the worldvolume of a stack on $N$ M5-branes probing a $\mathbb{C}^2/\Gamma_G$ orbifold \cite{DelZotto:2014hpa}, where $\Gamma_G$ is the finite simple group corresponding to the same ADE-type as $G$. 

Since all the necessary building blocks have 6d realizations, we expect there to be such descriptions for the entire set of theories we consider in this paper. (See \cite{Xie:2016uqq, Xie:2017vaf, Xie:2017aqx} for examples.) A six-dimensional description would also provide the Seiberg--Witten geometry. Furthermore, it would be interesting to see which $\widehat{\Gamma}(G)$ theories can have both 6d $(1,0)$ and $(2,0)$ origins, especially because a 6d $(1,0)$ SCFT origin may provide a geometric construction from the F-theory perspective \cite{Heckman:2013pva,Heckman:2015bfa,Morrison:2016djb}.\footnote{The geometric engineering of F-theory via elliptic fibrations provides different geometric perspectives for 6d $(1,0)$ theories (see \cite{Esole:2017rgz,Esole:2017qeh,Esole:2017hlw,Esole:2018csl,Esole:2018mqb,Esole:2019asj,Esole:2015xfa,Esole:2014bka,Esole:2014hya,Esole:2019hgr,Esole:2019rzq} for some explicit geometric constructions) and we get a superconformal field theory when it is compactified on a non-compact Calabi--Yau threefold satisfying certain conditions. The geometric construction for 6d $(1,0)$ SCFTs has shown to be a powerful approach: see \cite{DelZotto:2018tcj,Heckman:2018pqx,Cabrera:2019izd,Apruzzi:2019vpe,Razamat:2019mdt,Apruzzi:2019opn,Hassler:2019eso,Apruzzi:2019enx,Baume:2020ure,Baume:2021qho} for recent samples of indicative applications.}

Another interesting aspect of $\widehat{\Gamma}(G)$ theory is the associated vertex operator algebra (VOA) or chiral algebra \cite{Beem:2013sza}. Among them, the associated chiral algebra for the $\widehat{E}_6(SU(2))$ theory has already been explored in detail in \cite{Buican:2016arp}, where it is conjectured to be the $\mathcal{A}(6)$ algebra of Feigin--Feigin--Tipunin \cite{Feigin:2007sp,MR2766985}. 
It is straight-forward to construct the VOA corresponding to $\widehat{\Gamma}(G)$ theories without flavor symmetry, since the VOA for $\mathcal{D}_p(G)$ is given in terms of an affine Lie algebra \cite{Xie:2016evu, Song:2017oew}. One can then simply gauge the flavor symmetry $G$ to construct the VOA for the $\widehat{\Gamma}(G)$. However, it is desirable to have more intrinsic definition without referring to gauging. 

As we have seen already, it is possible to compute the Schur index of $\widehat{\Gamma}(G)$ theory if it does not have any flavor symmetry. A useful way of computing the Schur index is to use 4d/2d correspondence that maps the index of class $\mathcal{S}$ theory to the correlation function of 2d topological field theory \cite{Gadde:2009kb, Gadde:2011ik, Gadde:2011uv}, which has been further extended to the Argyres--Douglas theories \cite{Buican:2015ina, Buican:2015tda, Song:2015wta, Buican:2017uka, Watanabe:2019ssf}. For some cases, the Macdonald index, which is slightly more refined then the Schur index, can be computed in this way. It would be interesting to improve our result by further computing the Macdonald index as was done in \cite{Buican:2015tda, Song:2015wta, Watanabe:2019ssf} beyond $G=SU(N)$ case. One can also obtain the Macdonald index from the associated VOA \cite{Song:2016yfd, Agarwal:2018zqi, Agarwal:2021oyl, Foda:2019guo, Watanabe:2019ssf, Beem:2019tfp,Xie:2019zlb}, which may clarify the connection between $\mathcal{N}=4$ SYM theory and the $\widehat{\Gamma}(G)$ theories.  

The $\widehat{\Gamma}(G)$ theories we have constructed and explored may be further expanded to more broad set of theories and give various applications to one-form symmetries and Nekrasov partition functions. Let us make some comments about these additional points for future research directions.
\begin{itemize}
    \item Extension to BCFG: We have considered gauging together $\mathcal{D}_p(G)$ theories with $G$ an $ADE$ Lie group. The $\mathcal{D}_p(G)$ theories also exist when $G$ is non-simply-laced \cite{Wang:2018gvb,Carta:2021whq}, and a similar gauging procedure leads to an interesting additional set of theories, both with $a = c$, and beyond. 

    \item Extension to $\mathcal{D}_p^{b}(G)$ with $b \neq h_G^\vee$: In the current paper, we have only considered the case with $b=h_G^\vee$ which is not the most general AD theory with $G$ flavor symmetry. It would be interesting to consider the $b \neq h_G^\vee$ theories as additional building blocks as well, where some cases had been analyzed in \cite{Closset:2020afy}. The Schur indices for such theories are not available yet. 
    
    \item Higher-form symmetries: The $\widehat{\Gamma}(G)$ theory has a 1-form symmetry given by the center of $G$ \cite{Gaiotto:2014kfa, Aharony:2013hda}. This allows us to consider gauging by $G$ quotiented by a subgroup of the center. This affects the line operator spectrum and dualities.  It would be interesting to understand the consequences of such symmetries in our setup. One-form symmetry for the AD theory has been studied recently in \cite{Closset:2020scj,DelZotto:2020esg,Buican:2021xhs}. 

    \item Nekrasov partition function: The $\mathcal{D}_p(G)$ theory is generically non-Lagrangian and there is no direct way of computing the Nekrasov partition function for non-Lagrangian theories. However, for some special cases that involve gauging of Argyres--Douglas theories, there exists a way of computing the Nekrasov partition function through the AGT correspondence \cite{Alday:2009aq, Gaiotto:2009ma}, which has been done for some cases in \cite{Bonelli:2011aa, Gaiotto:2012sf, Kimura:2020krd}. It would be interesting to look for a further generalization of this analysis to $\widehat{\Gamma}(G)$ theories, which involve gauged Argyres--Douglas theories. 
    
\end{itemize}
As we have discussed thus far, we have just scratched the surface of the rich structure of the $\widehat{\Gamma}(G)$ theories. We hope to return to answer some of the points we have raised here in the near future. 

\subsection*{Acknowledgements}

We thank Prarit Agarwal for providing us with an extensive database of Argyres--Douglas theories. We also thank Seungkyu Kim for helping us compute the Schur indices for higher-rank theories to high orders. 
C.L.~was supported by a University Research Foundation grant at the University of Pennsylvania and DOE (HEP) Award DE-SC0021484.
M.J.K.~and J.S.~are partly supported by the National Research Foundation of Korea (NRF) grant NRF-2020R1C1C1007591. M.J.K.~is also supported by NRF-2020R1A4A3079707, a Sherman Fairchild Postdoctoral Fellowship, and the U.S.~Department of Energy, Office of Science, Office of High Energy Physics, under Award Number DE-SC0011632. J.S.~is also supported by the Start-up Research Grant for new faculty provided by Korea Advanced Institute of Science and Technology (KAIST).

\appendix

\section{Central charge coefficients}

In this appendix, we include a table of the coefficients $b(\widehat{\Gamma}(G))$ that appears in the formula for the central charges $a(\widehat{\Gamma}(G))$ in equation \eqref{eqn:centralcharges}. These are written in Table \ref{tbl:b}. This information is the same as that incorporated in Table \ref{tbl:bofDE}, however we find that the format in Table \ref{tbl:b} is occasionally easier to work with, and we include it here for future convenience.

\begin{table}[H]
\begin{threeparttable}
    \centering
    \small
    \renewcommand{\arraystretch}{1.3}
    $\begin{array}{c|c|c|c|c}
    \toprule
        \phantom{GGroup}G\phantom{GGroup} & 
        \phantom{\widehat{D}_4}b(\widehat{D}_4(G))\phantom{\widehat{D}_4} & \phantom{\widehat{D}_4}b(\widehat{E}_6(G))\phantom{\widehat{D}_4} & \phantom{\widehat{D}_4}b(\widehat{E}_7(G))\phantom{\widehat{D}_4} & \phantom{\widehat{D}_4}b(\widehat{E}_8(G))\phantom{\widehat{D}_4} \\\midrule
        SU(12\ell - 10) & \frac{1}{8} & 0 & \frac{1}{16} & \frac{1}{24} \\
        SU(12\ell - 9) & 0 & \frac{1}{6} & 0 & \frac{1}{12} \\
        SU(12\ell - 8) & \frac{1}{8} & 0 & \frac{3}{16} & \frac{1}{24} \\
        SU(12\ell - 7) & 0 & 0 & 0 & 0 \\
        SU(12\ell - 6) & \frac{1}{8} & \frac{1}{6} & \frac{1}{16} & \frac{5}{24} \\
        SU(12\ell - 5) & 0 & 0 & 0 & 0 \\
        SU(12\ell - 4) & \frac{1}{8} & 0 & \frac{3}{16} & \frac{1}{24} \\
        SU(12\ell - 3) & 0 & \frac{1}{6} & 0 & \frac{1}{12} \\
        SU(12\ell - 2) & \frac{1}{8} & 0 & \frac{1}{16} & \frac{1}{24} \\
        SU(12\ell - 1) & 0 & 0 & 0 & 0 \\
        SU(12\ell) & \frac{1}{8} & \frac{1}{6} & \frac{3}{16} & \frac{5}{24} \\
        SU(12\ell + 1) & 0 & 0 & 0 & 0 \\\midrule
        SO(24\ell - 16) & \frac{3\ell}{2} - 1 & -\frac{1}{6} & \frac{3\ell}{4} - \frac{1}{2} & \frac{\ell}{2} - \frac{1}{3} \\
        SO(24\ell - 14) & \frac{3\ell}{2} - \frac{9}{8} & 0 & \frac{3\ell}{4} - \frac{11}{16} & \frac{\ell}{2} - \frac{3}{8} \\
        SO(24\ell - 12) & \frac{3\ell}{2} - \frac{3}{4} & 0 & \frac{3\ell}{4} - \frac{3}{8} & \frac{\ell}{2} - \frac{1}{4} \\
        SO(24\ell - 10) &  \frac{3\ell}{2} - \frac{7}{8} & -\frac{1}{6} & \frac{3\ell}{4} - \frac{5}{16} & \frac{\ell}{2} - \frac{11}{24} \\
        SO(24\ell - 8) & \frac{3\ell}{2} - \frac{1}{2} & 0 & \frac{3\ell}{4} - \frac{1}{4} & \frac{\ell}{2} - \frac{1}{6} \\
        SO(24\ell - 6) & \frac{3\ell}{2} - \frac{5}{8} & 0 & \frac{3\ell}{4} - \frac{7}{16} & \frac{\ell}{2} - \frac{5}{24} \\
        SO(24\ell - 4) & \frac{3\ell}{2} - \frac{1}{4} & -\frac{1}{6} &\frac{3\ell}{4} - \frac{1}{8} & \frac{\ell}{2} - \frac{1}{12} \\
        SO(24\ell - 2) & \frac{3\ell}{2} - \frac{3}{8} & 0 & \frac{3\ell}{4} - \frac{1}{16} & \frac{\ell}{2} - \frac{1}{8} \\
        SO(24\ell) & \frac{3\ell}{2} & 0 & \frac{3\ell}{4} & \frac{\ell}{2} \\
        SO(24\ell + 2) & \frac{3\ell}{2} - \frac{1}{8} & -\frac{1}{6} & \frac{3\ell}{4} - \frac{3}{16} & \frac{\ell}{2} - \frac{5}{24} \\
        SO(24\ell + 4) & \frac{3\ell}{2} + \frac{1}{4} & 0 & \frac{3\ell}{4} + \frac{1}{8} & \frac{\ell}{2} + \frac{1}{12} \\
        SO(24\ell + 6) & \frac{3\ell}{2} + \frac{1}{8} & 0 & \frac{3\ell}{4} + \frac{3}{16} & \frac{\ell}{2} + \frac{1}{24} \\\midrule
        E_6 & \frac{1}{4} & \frac{1}{2} & -\frac{1}{8} & \frac{1}{4} \\\midrule
        E_7 & \frac{7}{8} & \frac{1}{3} & \frac{7}{16} & \frac{7}{24} \\\midrule
        E_8 & 1 & \frac{2}{3} & \frac{1}{2} & \frac{1}{3} \\\bottomrule
    \end{array}$
\end{threeparttable}
\caption{The $b(\widehat{\Gamma}(G))$ coefficients in the central charge $a(\widehat{\Gamma}(G))$ in equation \eqref{eqn:centralcharges}.}
    \label{tbl:b}
\end{table}

\bibliography{references}{}
\bibliographystyle{sortedbutpretty}
\end{document}